\documentclass[aps,prb,preprint,superscriptaddress,showpacs,onecolumn]{revtex4}

\usepackage{gensymb}
\usepackage{amsmath}
\usepackage{graphicx}
\usepackage{natbib}
\usepackage{amssymb}
\usepackage{amsthm}
\usepackage{braket}
\usepackage{caption}
\usepackage[subrefformat=parens,labelformat=parens]{subcaption}

\usepackage{bbding,pifont}

\newcommand{\figref}[1]{Fig.~\ref{#1}} % For not at the start of a sentence
\newcommand{\secref}[1]{Sec.~\ref{#1}} % For not at the start of a sentence
\newcommand{\appref}[1]{Appendix~\ref{#1}} 
\newcommand{\figureref}[1]{Figure~\ref{#1}} % For at the start of sentence
\newcommand{\tabref}[1]{Table~\ref{#1}}

\usepackage{xcolor,xspace}
\newcommand{\ds}{\displaystyle}

\newcommand{\pounders}{POUNDerS\xspace}

\usepackage[normalem]{ulem} % For \sout
 
\begin{document}

%Title of paper
%
\title{Origins and optimization of entanglement in plasmonically coupled quantum dots}

% Authors / affiliations
\author{Matthew Otten}
\affiliation{Department of Physics, Cornell University, Ithaca, NY 14853}
\affiliation{Mathematics and Computer Science Division, Argonne National Laboratory, Lemont, IL 60439}

\author{Jeffrey Larson}
\affiliation{Mathematics and Computer Science Division, Argonne National Laboratory, Lemont, IL 60439}

\author{Misun Min}
\affiliation{Mathematics and Computer Science Division, Argonne National Laboratory, Lemont, IL 60439}

\author{Stefan M.~Wild}
\affiliation{Mathematics and Computer Science Division, Argonne National Laboratory, Lemont, IL 60439}

\author{Matthew Pelton}
\affiliation{Department of Physics, University of Maryland, Baltimore County, Baltimore, MD 21250}

\author{Stephen K.~Gray}
\affiliation{Center for Nanoscale Materials, Argonne National Laboratory, Lemont, IL 60439}

% Date
\date{\today}

% Abstract
\begin{abstract}

A system of two or more quantum dots interacting with a dissipative
plasmonic nanostructure is investigated in detail by using a 
cavity
quantum electrodynamics approach with a model Hamiltonian.
We focus on determining and understanding system configurations that 
generate multiple bipartite quantum entanglements between the
occupation states of the quantum dots. 
These configurations include allowing for the quantum dots to
be asymmetrically coupled to the plasmonic system.
Analytical solution of a simplified limit for an arbitrary number of
quantum dots 
and numerical
simulations and  optimization for the two- and three-dot
cases are used to develop guidelines for maximizing the
bipartite entanglements. 
For any number of quantum dots, we show that through simple starting
states and parameter guidelines, one quantum dot can be made to share
a strong amount of bipartite entanglement with all other quantum dots in
the system, while entangling all other pairs to a lesser degree.
\end{abstract}

% PACS Numbers

% insert suggested keywords - APS authors don't need to do this
%\keywords{}

%\maketitle must follow title, authors, abstract, \pacs, and \keywords
\maketitle

\section{Introduction}

Hybrid systems composed of plasmonic nanostructures and quantum 
emitters/absorbers  (e.g., semiconductor quantum dots 
or other gain media) are currently of much interest.   This
is due to the plasmonic
element's ability to interact strongly with light, leading to subwavelength
control and confinement, and the possibility of the quantum element 
introducing nonlinear optical (as well as more general quantum) responses 
that may be enhanced by its proximity to the
plasmonic element.  
For example, theoretical predictions of certain 
Fano resonance phenomena~\cite{sasha_prl,savasta_cqed,shah-prb-2013} in such
systems exist, and experiments
are approaching the ability to measure such features~\cite{elaine-li-2015a}
and 
have already demonstrated
novel lasing action~\cite{odom-2013} and quantum coherences.~\cite{elaine-li-2015b}
The studies of such hybrid systems can also be viewed 
as steps in the emerging field of
``quantum plasmonics,''~\cite{tame-natphys-2013,fakonas-natphot-2014,
thakkar-acsphot-2015}
which aims to realize quantum-controlled devices relevant to
quantum sensing, single-photon sources, and nanoscale electronics.
 
% from old prb:
%Plasmonic nanostructures provide the potential for extremely strong
%light-matter interactions because of their ability to concentrate optical
%fields to nanometer-scale dimensions.~\cite{pelton-book-2013}
%Such strong interactions could
%enable the manipulation of quantum
%states in materials that interact with the 
%confined fields.~\cite{Lin2010,Chen2011,Chang2014}  However, plasmon resonances
%are necessarily associated with strong dissipation, raising the question
%of whether quantum effects are compatible with such rapid loss
%of energy and coherence.

% This is a modified form of our old second paragraph -- may wish to
% further modify!
Features relevant to quantum
information, such as the entanglement among the quantum dots (QDs), 
can also be achieved in hybrid plasmonic/QD systems.
This ability may seem surprising given the dissipative
(or lossy) aspect of plasmonic structures.  However, 
interactions between quantum objects and a dissipative environment
lead to the 
production of stable entangled states.~\cite{Poyatos1996,Lin2013,Krauter2011}
Several pioneering theoretical
studies have shown that dissipation-induced
entanglement is relevant to systems of QDs interacting
with plasmonic nanostructures.~\cite{vidal-prl-2011,vidal-prb-2011,
he-nanoscale-2012,he-quantinfo-2013,Gonzalez2013,Lee2013,
artuso-bryant-prb-2013}

We previously explored methods for generating entanglement in QD-plasmon systems,
using both systems in which only one QD is initially prepared in its excited state
and the system evolves without external excitation
and systems in which all the QDs are initially in their ground states and 
the entire system is excited by an ultrafast laser pulse.~\cite{otten-prb-2015} 
We showed that either a single or 
repeated optical pulse entangled the QDs and that the amount of 
entanglement can be tuned by controlling the coupling of the QDs to the 
plasmonic nanostructure. Furthermore, the whole system can be excited with a single
pulse, without the need to individually address each subsystem. 
This work and the present work allow for asymmetric coupling of the QDs to the
plasmonic system; for example, one can imagine the QDs to be configured to be at
different distances from the plasmonic system or in some other way that can
lead to asymmetry.

To allow dissipation-induced
entanglement to be an effective candidate for quantum information applications, 
one must thoroughly
understand how the entanglement is generated. Furthermore, constraining 
parameter sets in experimentally viable regions of the parameter space and knowing the sets'
associated degrees of entanglement are important for engineering such
systems within any larger quantum information platform. 
In this paper, we seek to determine system features that  
maximize the degree of entanglement between the QDs. To 
accomplish this objective, we employ analysis based on 
solutions of limiting forms
of the problem and
optimization based on numerical solutions to the complete 
cavity quantum electrodynamics equations. We show that for any number of QDs, simple
initial conditions and parameter guidelines generate systems where 
all pairs of QDs share some degree of entanglement.

\section{Theoretical Methods}\label{sec:theoretical_methods}
% \subsection{System} % SW: No need for a subsection here
We consider a cavity quantum electrodynamics (CQED) 
model of a system of $N$ quantum dots in proximity to a plasmonic system.
The underlying system's basis states are 
\begin{equation}\label{primitive}
\Ket{q_{N}, q_{N-1}, \ldots , q_1; s} = \ket{q_N} \ket{q_{N-1}} \ldots \ket{q_1} 
\ket{s},
\end{equation}
%%%%%%%%%%%
% NOTE: "i" is index, but q_i and s itself is not indices. So... I'll try the following.
%%%%%%%%%%%
%where $q_i \in \set{0,1} $ indexes the exciton of the $i$th QD and
%$s \in \set{0, 1, 2, \ldots, N_s} $ indexes the plasmon energy levels.  
where $q_i \in \set{0,1} $ represents the exciton of the $i$th QD and
$s \in \set{0, 1, 2, \ldots, N_s} $ represents the plasmon energy levels.  
%%%%%%%%%%%
%NOTE: Stephen, should we use just \sum_{q} and \sum_{q'} without s and s' below?
%      in case if we want to distinguish the summation of s in (8)?
%      My understanding was that, in (8), we really do summation for all the possible "s".     
%      But in (2), we have only one term for s. So... please correct any inaccuracy 
%      that I might have made.
%%%%%%%%%%%
Using a simplified notation $q= {q_{N},\ldots,q_1}$, we can write the density operator as
\begin{equation}\label{DM}
  \begin{array}{rl}
    \hat{\rho}(t) = \ds \sum_{qs} & \ds \sum_{q's'} C_{qs,q's'}(t) \ket{q;s}\bra{ q';s'}.
 \end{array}
\end{equation}
%%%%%%%%%%%%
%NOTE: If above is not prefered, please feel free to change back to original below.
%Our density operator may thus be written as
%\begin{equation}\label{DM}
%  \begin{array}{rl}
%   \hat{\rho}(t) = \ds \sum_{q_{N},\ldots,q_1;s} & \ds \sum_{q'_{N},\ldots,q'_1;s'} 
%                       \Big[C_{q_{N},\ldots,q_1;s,~q'_{N},\ldots,q'_1;s'}(t)\\
%    &\ket{q_{N},\ldots,q_1;s}\bra{ q'_{N},\ldots,q'_1;s'}\Big]
%  \end{array}
%\end{equation}
%%%%%%%%%%%%
%NOTE: giving some weight on the governing equation using a separate sentence.
Then our governing equation describing the CQED model is defined as 
\begin{equation}\label{master_equation}
\frac{\mathrm{d} \hat{\rho}}{\mathrm{d} t}
= - \frac{i}{\hbar} [ \hat{H}, \hat{\rho} ] 
 - \frac{i}{\hbar} [ \hat{H_d}, \hat{\rho} ] +
L(\hat{\rho}),
\end{equation}
where $\hat{H}$, $\hat{H_d}$, and $L$ are the operators for the Hamiltonian, the driving term, 
and the Lindblad, respectively.
The Hamiltonian $\hat{H}$ for the coupled dot-plasmon system as
\begin{equation}\label{hamiltonian}
\hat{H} = \sum_i \hat{H}_i + \hat{H}_s + \sum_i \hat{H}_{s,i}.
\end{equation}
Defining the lowering and raising operator pairs for both the QDs and the plasmon, 
$(\hat{\sigma}_i, \hat{\sigma}_i^\dagger)$ and $(\hat{b},\hat{b}^\dagger)$, in the usual manner 
as in~\cite{shah-prb-2013,otten-prb-2015}, we have the isolated dot and plasmon Hamiltonian terms 
\begin{equation}
  \hat{H}_i = \hbar \omega_i \hat{\sigma}_i^{\dagger} \hat{\sigma}_i \qquad \mbox{and} \qquad
 \hat{H}_s = \hbar \omega_s \hat{b}^{\dagger} \hat{b}, 
\end{equation}
respectively, and the dot-plasmon coupling terms 
\begin{equation}\label{g_term}
  \hat{H}_{s,i} = - \hbar g_{i} ( \hat{\sigma}_i^{\dagger} \hat{b} + 
  \hat{\sigma}_i \hat{b}^{\dagger} ).
\end{equation} 
Equation \eqref{g_term} represents the simplest possible dot-plasmon
coupling term corresponding to a QD gaining (losing) a quantum
of energy when the plasmon loses (gains) a quantum of energy.

For the system exposed to a time-dependent electric field $E(t)$, 
we have %the driving term as %is defined as %in \eqref{master_equation} is
\begin{equation}
\hat{H}_d = - E(t) \left[
 \sum_i d_i ( {\hat{\sigma}_i} + \hat{\sigma}_i^{\dagger} ) + 
 d_s ( \hat{b} + \hat{b}^{\dagger} )  
\right],
\end{equation}
where $ d_i $ and $ d_s $ denote the
transition dipole moments of the QDs and 
plasmon, respectively.

We assume that the distance between the QDs is large compared with the separation
between QDs and neighboring metal nanoparticles, so that direct through-space
coupling among the QDs can be neglected.
We also neglect retardation; hence, our treatment is limited
to systems with physical dimensions that are small compared with optical wavelengths.

The Lindblad superoperator $ L(\hat{\rho}) $ in \eqref{master_equation}  
describes the
dephasing and dissipation effects. 
We employ a previously developed~\cite{shah-prb-2013} extension of
$L(\hat{\rho})$ that is parameterized by the QD population 
decay $\gamma_p$, the QD dephasing rate $\gamma_{d}$, and the
plasmon decay constant $\gamma_s$. 
We consider time scales on the order of the inverse of these rates,
so that there are no correlated fluctuations in the QDs' states and so that the 
use of the Lindblad superoperator is justified. 
Although environmental dephasing is explicitly included for the
QDs, it is not necessary to do so for the plasmon because the dephasing that 
arises from its decay (encoded in the corresponding term in $L(\hat{\rho})$)
is much larger in magnitude.
As in Ref.~\onlinecite{shah-prb-2013}, the rotating wave approximation is 
applied.

We use Wootter's concurrence~\cite{wootters-prl-1998} to measure the entanglement of the system.
%NOTE: The density matrix description \rho is mentioned before the reduced matrix \rho'
An alternative representation of the density operator \eqref{DM} is the density matrix $\rho$ 
with its elements defined by
\begin{equation}\label{density}        
\rho_{qs,q' s'}=\bra{q;s}\hat\rho \ket{q' ; s'}.
\end{equation}                       
Let $\rho'$ be the reduced density matrix associated with one particular pair of QDs, $A$ and $B$, 
obtained by tracing the full density matrix $\rho$ over the 
plasmon quantum numbers $s$ and the quantum
numbers for all other QDs. 
The $AB$ pairwise concurrence is then given by
\begin{equation}\label{concurrence}
C_{A,B} = \max\{0, \lambda_1 - \lambda_2 - \lambda_3 - \lambda_4\}, 
\end{equation}
where $\lambda_i$ are the square roots of the eigenvalues 
of $\rho'\tilde{\rho}'$ with $\lambda_i\geq \lambda_{i+1}$ (in descending order). 
The matrix  $\tilde{\rho}'$ corresponds to
the spin-flipped density matrix~\cite{wootters-prl-1998}
\begin{equation}
\tilde{\rho}' = (\sigma_y \otimes \sigma_y)  (\rho')^* (\sigma_y \otimes \sigma_y),
\end{equation}
where
\begin{equation}
(\sigma_y \otimes \sigma_y) = 
\begin{bmatrix}
0  & 0 & 0 & -1   \\
0  & 0 & 1 & 0    \\
0  & 1 & 0 & 0    \\
-1 & 0 &0  & 0   \\
\end{bmatrix}.
\end{equation}
%%%%
% NOTE: I'm just removing 4x4 size matrix eigenvalue computation discussion.      
%%%%

\subsection{Approximate Analysis}
\label{analytical}
We define the ``dark'' evolution to be how a QD-plasmon system evolves given
some initial QD excitation with everything else in the system initially in the ground state.  
In the limit of
low total excitation 
energy one can develop an exact 
analytical 
solution
for the problem of an arbitrary number of quantum dots interacting with
the plasmon if QD dephasing is neglected. This procedure is discussed in \appref{three-state-model}.
We first discuss some predictions from this analysis
for two QDs and then briefly for larger numbers of QDs.  
Also of interest is the case of pulsed excitation, where
the system is initially cold and then subjected to a laser pulse.
We follow the discussion of the dark evolution
with analysis
of this pulsed case using simple Rabi flopping ideas.

In the case of two QDs coupled to a plasmon, we
are concerned with determining QD-plasmon coupling factors ($g_1$ and
$g_2$) that maximize entanglement.  
For two QDs in particular, it is convenient to consider the 
two entangled QD states
\begin{equation}\label{sym_state}
\ket{S;s} = \frac{1}{\sqrt{2}} \Big[ \ket{q_2=0}\ket{q_1=1} 
+ \ket{q_2=1}\ket{q_1=0} \Big] \ket{s}
\end{equation}
and
\begin{equation}\label{asym_state}
\ket{A;s} = \frac{1}{\sqrt{2}} \Big[ \ket{q_2=0}\ket{q_1=1} 
- \ket{q_2=1}\ket{q_1=0} \Big] \ket{s}
\end{equation}
in our calculations instead of the direct product of primitive QD states as in 
\eqref{primitive}.
For two QDs, 
\appref{three-state-model} 
discusses in detail 
a three-state Hamiltonian model involving the 
basis states 
$\ket{q_2=0,q_1=0;s=1}$, $\ket{S;s=0}$, and
$ \ket{A;s=0}$ that neglects QD dephasing and spontaneous emission but 
allows for plasmon dissipation by introducing an
appropriate complex diagonal matrix element to the Hamiltonian
matrix.
The initial state of relevance to the dark limit calculations is one
with an excited QD1, an unexcited QD2, and a plasmon;
this state is represented by 
$\ket{0,1;0} = \frac{1}{\sqrt{2}} ( \ket{S;0} + \ket{A;0} )$.
This initial state is interesting because although it is a separable,
unentangled state, it is a nonstationary state 
of the full system
that has been shown to evolve 
into a state with a possibly significant 
transient degree of entanglement.~\cite{vidal-prl-2011,vidal-prb-2011,otten-prb-2015}
With no plasmon dissipation ($\gamma_s = 0$) 
and for short times $t$,  the 
probabilities of states
$\ket{0,0;1}$, $\ket{S;0}$, and $\ket{A;0}$ are given by the respective 
squares of $a_0(t)$, $a_S(t)$, and $a_A(t)$, where 
\begin{equation}\label{approx_g2}
  \begin{bmatrix}
    a_0(t)  \\
    a_S(t)  \\
    a_A(t) 
  \end{bmatrix}  
 \approx 
  \begin{bmatrix}
    -i g_1 t   \\
    \frac{1}{\sqrt{2}}-\frac{g_1 (g_1+g_2 )t^2}{2\sqrt{2}}  \\
    \frac{1}{\sqrt{2}}+\frac{g_1 (g_2-g_1 )t^2}{2\sqrt{2}}
  \end{bmatrix}.
\end{equation}

When $\gamma_s > 0$, the limit as $t \rightarrow \infty$ is of interest
because the system can then reach a steady state in the populations. 
\appref{three-state-model} shows that for the initial condition with one excited QD and the
rest of the system unexcited,
\begin{equation}
  a_S ( \infty ) = \frac{1}{\sqrt{2} (1+x^2)} x (1 - x )
\end{equation}
and
\begin{equation}
  a_A ( \infty ) =  \frac{1}{\sqrt{2}(1+x^2)} (1 - x),
\end{equation}
where 
$x = \frac{g_1-g_2}{g_1+g_2}$.  Remarkably, these results are valid for any 
positive value of
$\gamma_s$ (although it must be remembered that no QD dephasing has been allowed).  The concurrence in this asymptotic limit is simply $|a_A(\infty )|^2 - | a_S(\infty )|^2$
and can be readily maximized to yield the optimum ratio of
coupling strengths:  $x = -2 + \sqrt{3}$  or 
\begin{equation}\label{opt_dark}
  \frac{g_2}{g_1} = \sqrt{3}. 
\end{equation}

\appref{three-state-model} also develops an {\em exact} procedure for constructing
the corresponding dark dynamics of $N$ QDs interacting with a plasmon,
without QD dephasing. This system 
is then described by an effective $(N+1) \times (N+1)$ complex effective Hamiltonian model.
For the scenario of one QD initially excited, it 
can be used to get an idea of how the entanglement results scale with increasing
$N$. 

We are also interested in the case when the system is initially unexcited and an optical
pulse is used to generate transient entanglement. Assuming the pulse
is relatively simple and resonant with the QDs' transition frequencies, a simple
question to ask is
what values of $g_1$ and $g_2$ will lead
to the two-QD system being close to the  
$\ket{0,1;0}$ state. We know from 
previous work~\cite{vidal-prl-2011,vidal-prb-2011,otten-prb-2015} that such a system 
will evolve 
into a state with some degree of entanglement. 
%Since we want to prepare the system in a state similar to the $\ket{0,1;0}$
%state, we can predict where the local maxima in a two-QD system will
%be. 
On resonance, the QDs undergo Rabi oscillations as 
they are excited by the
laser pulse.  
The time for QD$i$ to undergo one Rabi oscillation (i.e., to
go from the ground state to the excited state and then back to the ground
state) is $2 \pi / \Omega_R(i)$, where 
\begin{equation}\label{localref}
\Omega_R(i) = \frac{\mu_i E_0^{\rm{loc}}(i)}{\hbar} =
 \frac{2 g_i \mu_s E_0}{\hbar \gamma_s},
\end{equation}
with $E_0^{\rm{loc}}(i)$ being the amplitude of the sinusoidal electric field 
experienced locally by  
QD$i$. 
The final term in \eqref{localref} is obtained by using the expression 
for $E_0^{\rm{loc}}$
derived in \appref{local-field-enhancement}, which relates this local electric field 
to the incident field $E_0$.
(Other phenomena, such as the Purcell effect, are also
occurring; the Rabi formulae 
\eqref{localref} discussed here should be construed as 
approximate indicators of the dynamics.)
In order to achieve a highly entangled state, one QD, say QD1, must  
undergo $m-\frac{1}{2}$ Rabi oscillations (with $m = 1, 2, \ldots$) so that it is left in the excited
state. The time for this process to occur
is $2 \pi (m-\frac{1}{2}) / \Omega_R(1)$.   The other dot, QD2, must undergo $n = 1, 2, \ldots$ full 
oscillations so that it is left in its ground state.  
The time for this process to occur is 
$2 \pi n / \Omega_R(2)$.  
Equating these two Rabi times leads to the simple result that
\begin{equation}\label{opt_pulse}
  \frac{g_2}{g_1} = \frac{n}{m-\frac{1}{2}}.
\end{equation}

We see that the condition \eqref{opt_pulse} on the couplings for achieving one 
QD excited via pulsed excitation 
is not the same as the condition \eqref{opt_dark} on the couplings for that 
excited state
to evolve to an entangled state. 
Nonetheless,
for $m = n = 1$, \eqref{opt_pulse} predicts $g_2/g_1 = 2$,
in approximate accord with \eqref{opt_dark} where $g_2/g_1 \approx 1.73205$.
Although this restricts the parameter space for the pulsed case somewhat, many 
parameters (especially those describing the pulse) can still be varied freely.

\subsection{Concurrence Optimization}
To find the set of system parameters that maximize the sum of the pairwise concurrences, 
we employ a
numerical optimization framework that samples the parameter space 
in a
uniformly random fashion, evaluating the concurrence at each point. 
The parameters in question include the $N$ QD-plasmon coupling
coefficients ($g_i$, $i = 1, 2, \ldots, N$), environmental aspects
such as the QD dephasing and plasmon decay constants ($\gamma_d$ and
$\gamma_s$, respectively), and applied laser pulse features such as its
fluence ($F$) and duration ($\tau$). 
(See \secref{sec:sim} for definitions of the laser pulse parameters.)
Since the sum of the pairwise
concurrences is a nonconvex function of these parameters, 
several isolated local maxima are likely to exist.
Our approach follows that in Ref.~\onlinecite{larson-2015},
clustering evaluated points in the parameter space into basins of
attraction for different maxima. Clusters are formed by using the points' function
values (sum of pairwise concurrences) and their proximity to points with 
better function values. Points that do not have a better point
within a distance $d$ are considered the best points in their cluster. The 
distance $d$ can be adjusted so a reasonable number
of clusters are identified. (One also can dynamically
adjust $d$ as the parameter space is explored.~\cite{larson-2015,larson-2016})
Local optimization runs are then started from the best
point in each cluster.

The local optimization problem of maximizing the sum of pairwise 
concurrences 
is solved by minimizing the figure of merit 
\begin{equation}
\sum_{i<j} \left( 1-C_{i,j} \right)^2,
 \label{eq:obj}
\end{equation}
where $C_{i,j}$ depends on the system parameters being optimized over (see 
\eqref{concurrence}).
This form is appropriate because the pairwise concurrences in 
\eqref{eq:obj} are bounded above 
by 1. 
Depending on the context, $C_{i,j}$ might be the maximum concurrence
achieved over time or a long-time asymptotic limit.
When viewed as a function of the parameters, \eqref{eq:obj} 
defines 
a nonlinear least-squares problem.
We solve this problem with the Practical Optimization Using No
Derivatives for sums of Squares 
(\pounders) algorithm.~\cite{wild-2014,wild-2015}  
For a system with $N$ QDs, 
\pounders iteratively builds local quadratic
surrogates of each of the ${N \choose 2}$ residual functions $\left\{ 1 - 
C_{i,j} 
\right\}$ and
combines this information in a master surrogate model. In each iteration of 
the algorithm, this surrogate model is minimized
within a trust-region framework to generate candidate solutions. 

\subsection{Simulation Details}\label{sec:sim}
\label{numerical}
We consider the time evolution of the density operator in \eqref{master_equation},
with the choices of the parameters   
corresponding to a gold nanoparticle system interacting with QDs in a polymer matrix
with a dielectric constant $\epsilon_{med}$~=~2.25; these choices are similar
to those originally used in our single plasmonic-QD system study.~\cite{shah-prb-2013}
For QD$i$, we choose $\hbar \omega_i = \hbar \omega_s$~=~2.05~eV, assuming the QD and 
plasmon transition energies are equal. We set the QD dipole moments 
to be
$d_i$~=~13~D and the plasmon dipole moment to be 
$d_s$~=~4000~D. The QDs are assumed to have the same 
spontaneous decay rate, $\hbar \gamma_{p}$~=~190~neV. 
In some of our calculations we vary or consider several values
for the QD dephasing rate,
$\gamma_{d}$, and plasmon decay rate,
$\gamma_s$.  Consistent with
our earlier work, base values are 
 $\hbar \gamma_{d}$~=~2~meV and
$\hbar \gamma_s$~=~100~ meV. 
We utilize coupling factors, $\hbar g_i$, in the 0--30~meV range; 
and unlike all the other QD parameters,  we do allow QDs to have 
different coupling constant values.  
Previous calculations show that a realistic approximation for the plasmon-QD coupling is approximately 10~meV for 
a system such as the one we study here.~\cite{shah-prb-2013}
Other systems, such as silver nanoparticles or particles with different
geometries, could exhibit larger coupling factors than 
does gold.~\cite{wu-oe-2010}
For calculations that include a laser pulse $E(t)$, we assume it has
the form (in the nonrotating frame) $E(t) = G(t) E_0 \cos (\omega_0 t)$, where 
$\omega_0 = \omega_s = \omega_i$ and $G(t)$ is a 
Gaussian envelope function such that the full width at half maximum of
$E^2(t)$ is $\tau$.  The 
pulse fluence is $F = \int_{-\infty}^{+\infty} dt \sqrt \epsilon_{med} c \epsilon_0 
E^2(t)$.

We formulate a density matrix equation from \eqref{master_equation} using \eqref{density},
and we solve the density matrix equation consisting of a set of $M^2$ ordinary differential 
equations for the time-dependent complex amplitudes $C_{qs,q's'}(t)$, with $M = 2^N N_s$ where 
$N$ is the number of QDs and $N_s$ is the number of plasmon energy levels.
We solve these ODEs numerically using an efficient parallel solver that employs 
sparse matrix-matrix multiplication algorithms with either a Runge-Kutta 
or exponential time integration scheme.~\cite{misun_exp,kalu2013}

\section{Results}
% SW: Need a lead-off sentence.
We now detail our quantum dynamics results corresponding to a system of QDs
interacting with a plasmonic system as modeled in 
\secref{sec:theoretical_methods}. We analyze such systems for both free
evolution of some particular excitation (what we refer to as ``dark''
evolution) and in the presence of a laser pulse.

\subsection{Two Quantum Dots in the Dark}
\label{2-dots-dark}
We first consider two QDs (QD1 and QD2) interacting with a plasmonic
system under the assumption that the initial state 
$\ket{q_2=0,q_1=1; s=0}$ has been prepared and
% follow its subsequent time evolution 
evolves
in the absence of any external pulses,
that is, ``in the dark.'' 
Unlike cases studied in previous
work,~\cite{vidal-prl-2011,vidal-prb-2011,otten-prb-2015}
the possibility of asymmetric dot-plasmon couplings ($g_1 \neq g_2$) 
can lead to new features in the time-dependent concurrence. 

When the QDs are symmetrically distributed within
the plasmonic system so that $g_1 = g_2$, the $\ket{A;s}$ state 
is an eigenstate of the Hamiltonian \eqref{hamiltonian} and
decays with 
a relatively slow dephasing rate ($\gamma_d$) 
due only to the Lindblad term in \eqref{master_equation}.
With finite (but still symmetric) coupling,
the $|S;0\rangle$ state mixes with the $|S;1\rangle$ state~\cite{otten-prb-2015} and
is no longer an eigenstate of the Hamiltonian. The probability of being in $|S; 0\rangle$ 
is $\frac{1}{2} \cos^2(\sqrt{2}gt)$, leading to an increased initial decay of the 
$|S; 0\rangle$ state. 
The plasmon decay ($\gamma_s$) damps out any additional oscillations of the
$|S; 0\rangle$ population. 
As shown previously,~\cite{otten-prb-2015}
starting a system in the
$|0,1;0\rangle$ state then leads to a high degree of 
concurrence, since the
$|S;0\rangle$ state quickly decays, while the $|A;0\rangle$ state 
undergoes a much slower decay.
With $\gamma_d = 0$, the maximum concurrence for such a symmetric system 
is therefore
0.5. We describe here a method to achieve larger degrees of concurrence by 
forcing the 
$|S;0\rangle$ state to evolve into $|A;0\rangle$ rather 
than into $|0,0;1\rangle$.

By breaking the symmetry of the couplings between the two QDs, we mix the 
$|S; 0\rangle$ state with the $|A ;0\rangle$ state, through the 
$|0,0; 1\rangle$ state. This approach follows the 
analysis in \appref{three-state-model} where
a three-state model is discussed and solved analytically in certain
limits.
When $g_1=g_2$, no coupling occurs between
$|S; 0\rangle$ and $|A; 0\rangle$; but when $g_1 \ne g_2$, the two states
are indirectly coupled through the $|0,0; 1\rangle$ state (to which both states are directly
coupled). Starting in the 
$|0,1;0\rangle = (|A;0\rangle + |S;0\rangle)/2$ state, 
and setting $\hbar \gamma_s = \hbar \gamma_d$~=~0~meV,
lead to a cyclic evolution between a completely unentangled state and a highly 
entangled state. 

\begin{figure*}
  \centering
  \begin{subfigure}{0.5\textwidth}
    \centering
    \includegraphics[width=\columnwidth]{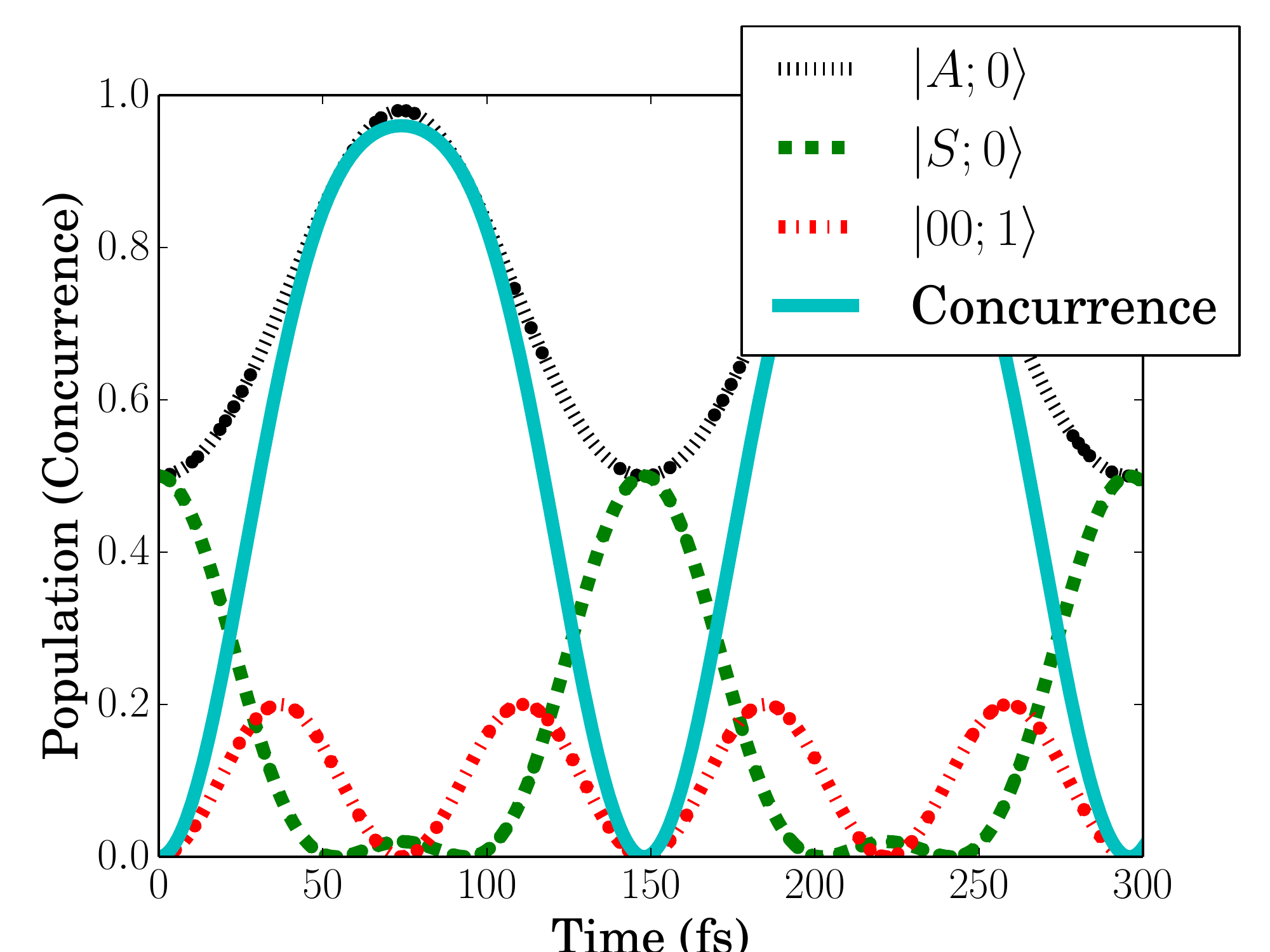}
    \caption{}\label{cyclic_evolution(a)}
  \end{subfigure}%
  ~
  \begin{subfigure}{0.5\textwidth}
    \centering
    \includegraphics[width=\columnwidth]{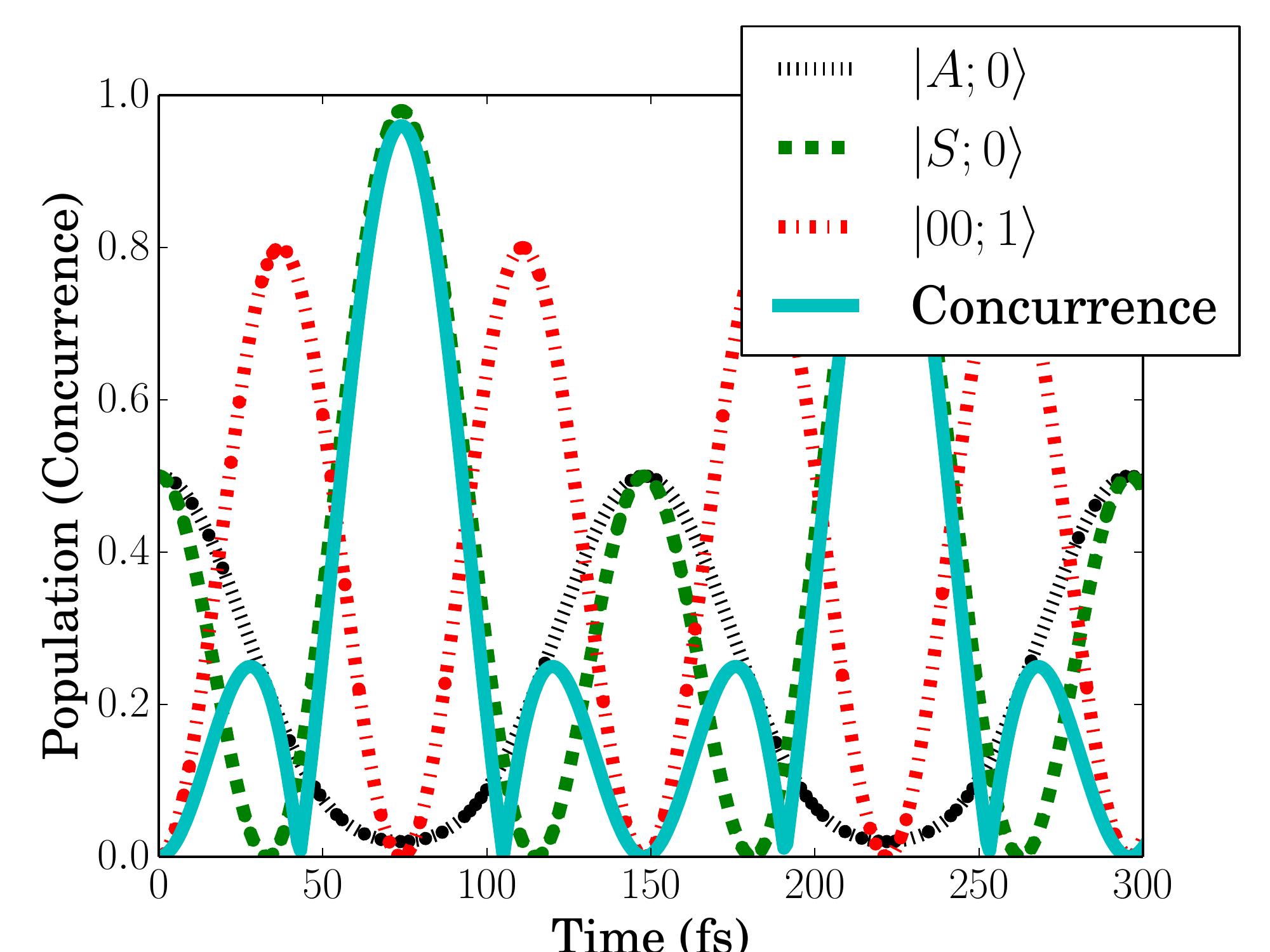}
    \caption{}\label{cyclic_evolution(b)}
  \end{subfigure}%
  \caption{
    Time-dependence of the populations of the 
states $|S;s=0\rangle$, $|A;s=0\rangle$, and $|q_2=0,q_1=0;s=1\rangle$, 
and the concurrence, for a two-QD system initially in the $|q_2=0,q_1=1;0\rangle$ state, 
with no surface plasmon decay or QD dephasing, i.e., 
$\hbar \gamma_s = \hbar \gamma_d$~=~0~meV. (a) A case with $g_1$ $<$ $g_2$ corresponding to 
the initially excited QD1 not being as strongly coupled to the surface
plasmon as QD2:  
    $\hbar g_1$~=~12.5~meV, $\hbar g_2$~=~25~meV.
(b) A case with $g_1$ $>$ $g_2$:
    $\hbar g_1$~=~25~meV, $\hbar g_2$~=~12.5~meV.
  } 
\label{cyclic_evolution}
\end{figure*}

We follow the convention that QD1 is the QD that is initially in its
excited state.
If 
$g_1 < g_2$, then the $|A; 0\rangle$ state reaches a population approaching 1 
(\figref{cyclic_evolution(a)});
 if $g_1 > g_2$, then the $|S; 0\rangle$ state reaches a population 
approaching 1 
(\figref{cyclic_evolution(b)}). 
The dynamics of these two examples are similar; the $|A; 0\rangle$
(resp.~$|S;0\rangle$) state evolves into the $|0,0;1\rangle$ state, which evolves
into the $|S;0\rangle$ (resp.~$|A;0\rangle$) state and back through the 
$|0,0;1\rangle$ state into its initial state. Where the 
$|S;0\rangle$ (resp.~$|A;0\rangle$) state reaches its maximum, the 
concurrence does as well, reaching a value of nearly 1.
Using the explicit three-state system described in \eqref{propagator0} in 
\appref{three-state-model}, 
we find the ratio $g_1/g_2 = \sqrt{2}-1 \approx 0.414$ gives
an $\ket{A;0}$ state population of unity, which also maximizes the concurrence for
the $g_1 < g_2$ case. A similar analysis can be done for the $\ket{S;0}$ state,
giving $g_1/g_2 = \sqrt{2}+1 \approx 2.414$.
Note that these optimal ratios for achieving large, instantaneous 
concurrences when there is no dissipation are different from 
those of \secref{analytical}. The latter concern
either an asymptotic concurrence that can be reached in the case of
dissipation or the couplings conducive to a pulsed laser generating
a particular excited state that can then evolve to 
a state with significant concurrence.

For short times $t$, the approximation \eqref{approx_g2}
applies. 
When $g_1 < g_2$ in this case, the second term of $a_A (t)$ is positive, which 
leads to a boost in the population of the $\ket{A}$ state. 
When $g_1 > g_2$, the second term is negative, and the 
population of the $\ket{A}$ state initially 
declines. In both cases,
$a_S (t)$ initially declines; but when $g_1 > g_2$, it reaches 0 much faster  
and then
rises to nearly 1. Both effects can be seen in 
\figref{cyclic_evolution}.

\begin{figure*}
  \centering
  \begin{subfigure}{0.5\textwidth}
    \centering
    \includegraphics[width=\columnwidth]{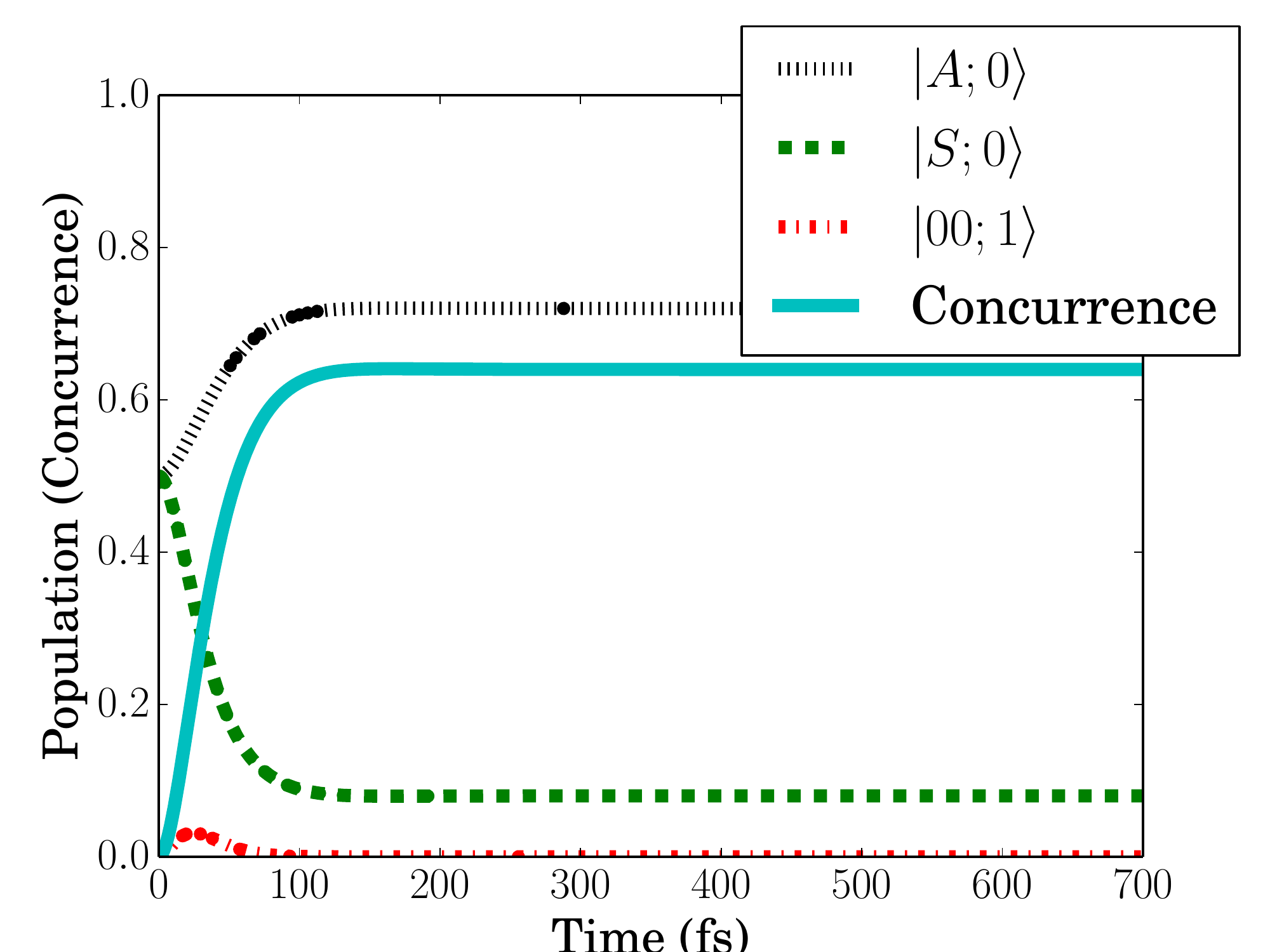}
    \caption{}\label{finite_gammas(a)}
  \end{subfigure}%
  ~
  \begin{subfigure}{0.5\textwidth}
    \centering
    \includegraphics[width=\columnwidth]{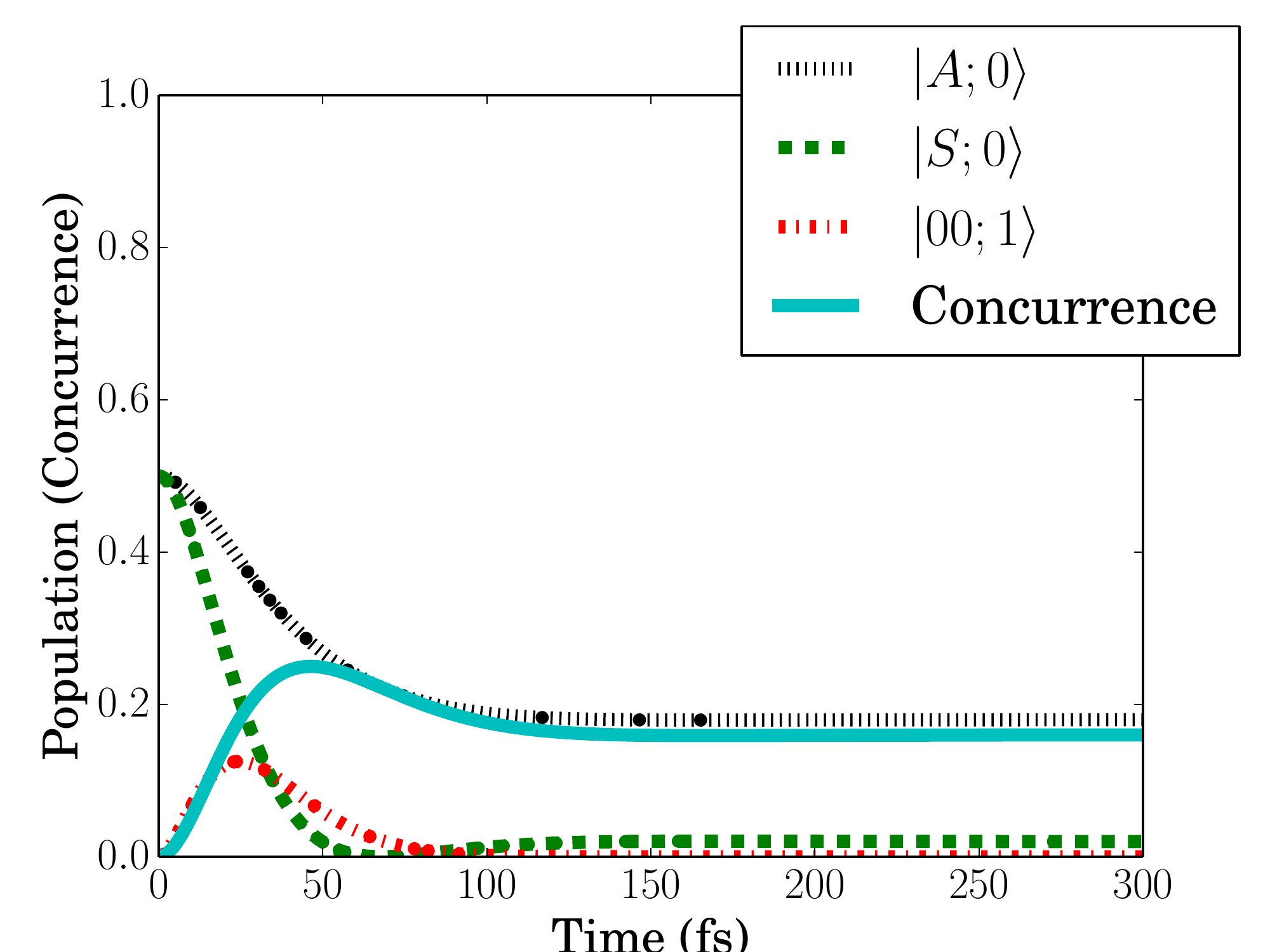}
    \caption{}\label{finite_gammas(b)}
  \end{subfigure}%
  \\
  \begin{subfigure}{0.5\textwidth}
    \centering
    \includegraphics[width=\columnwidth]{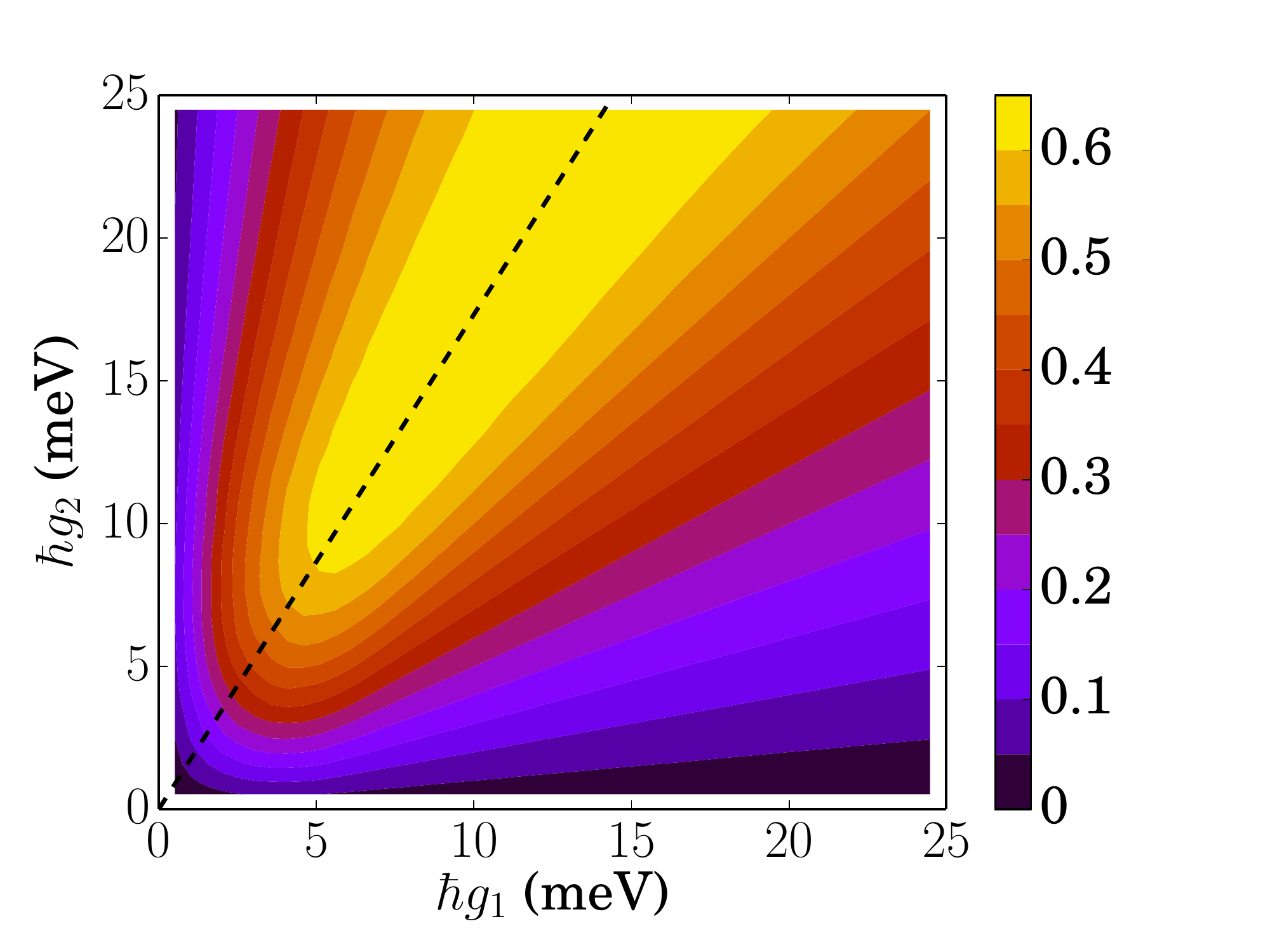}
    \caption{
      }\label{finite_gammas(c)}
  \end{subfigure}%
  \begin{subfigure}{0.5\textwidth}
    \centering
    \includegraphics[width=\columnwidth]{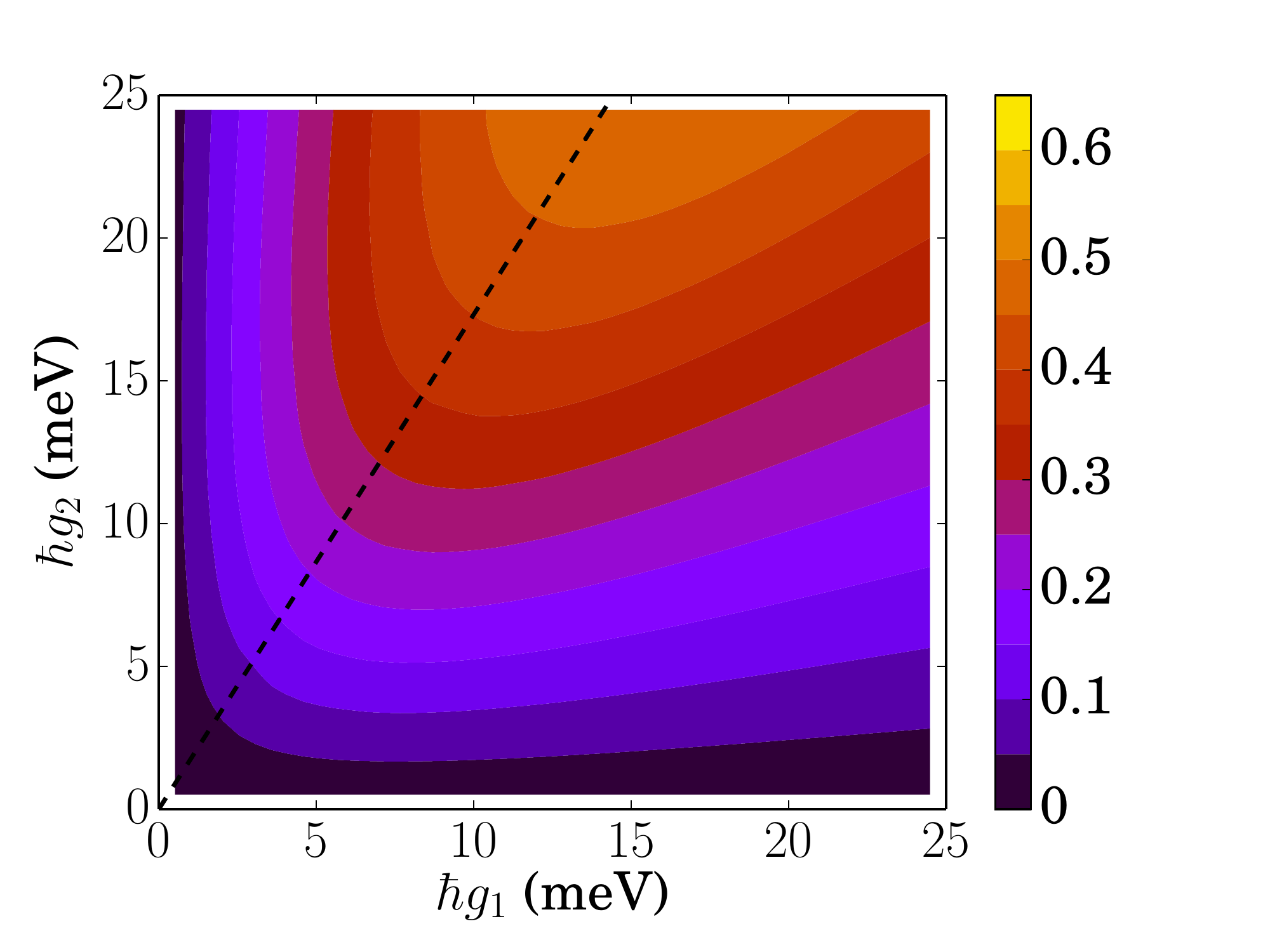}
    \caption{ 
}\label{finite_gammas(d)}
  \end{subfigure}%

  \caption{
Results for a surface plasmon decay width $\hbar \gamma_s$ = 100 meV.
Upper two panels (a) and (b) show the time-dependence of the
the $|S;s=0\rangle$, $|A;s=0\rangle$, $|q_2=0,q_1=0;s=1\rangle$ 
state populations 
and concurrence for a two-QD system, initially in the $|q_2=0,q_1=1;0\rangle$ state, 
and with no QD dephasing, i.e., $
\hbar \gamma_d$~=~0~meV. The cases with 
(a) 
    $\hbar g_1$~=~12.5~meV, $\hbar g_2$~=~25~meV
and 
(b) 
    $\hbar g_1$~=~25~meV, $\hbar g_2$~=~12.5~meV
 are displayed.  
The lower two panels (c) and (d)
are the maximum concurrences found as a function of the QD/plasmon
coupling factors, $g_1$ and $g_2$.  Panel (c) corresponds to
no QD dephasing, $\gamma_d$ = 0, and contains within it the 
concurrence maxima from the particular 
cases (a)
and (b) above. Panel (d) is the corresponding maximum concurrence when
the QD dephasing is set to  
$\hbar \gamma_d$~=~2.0~meV.
The dashed lines in (c) and (d) 
represent $g_2~$~=~$\sqrt{3}g_1$.
  } 
\label{finite_gammas}
\end{figure*}

When the Lindblad terms describing dissipation and dephasing are added, 
the results of the two simulations
($\hbar g_1,\hbar g_2$)~=~(12.5~meV, 25~meV) or (25~meV, 12.5~meV) in
\figref{cyclic_evolution} become very different. 
In \figref{finite_gammas} we consider simply adding 
plasmon dissipation ($\gamma_s > 0$) to the simulations, while keeping
the QD dephasing term $\gamma_d$ at zero. 
This case also has a closed-form solution (see \appref{three-state-model} and \secref{analytical}).
We see from Figs.~\ref{finite_gammas}\subref{finite_gammas(a)} and \ref{finite_gammas}\subref{finite_gammas(b)} that the  initial state evolves and begins to
populate the first excited plasmon state,  
but the plasmon population quickly decays and the system
reaches a steady state.
The steady-state concurrence for the
case with $g_1 < g_2$, Fig.~\ref{finite_gammas}\subref{finite_gammas(a)}, is larger than the case with
$g_1 > g_2$, Fig.~\ref{finite_gammas}\subref{finite_gammas(b)}.  
This trend might be expected on the
basis of the dynamics without plasmon decay,
Figs.~\ref{cyclic_evolution}\subref{cyclic_evolution(a)} and
\ref{cyclic_evolution}\subref{cyclic_evolution(b)},
wherein a smooth rise of concurrence from 0 to 1 occurs over initial
times for the case $g_1 < g_2$, Fig.~\ref{cyclic_evolution}\subref{cyclic_evolution(a)}, but a more complicated
behavior involving a small local maximum
in concurrence occurs for the case $g_1 > g_2$, ~\ref{cyclic_evolution}\subref{cyclic_evolution(b)}. 
In a realistic system, the $g_1 > g_2$ case will not
create concurrences as large as those seen when $g_1 < g_2$; the best case is 
that the plasmon decay is 
sufficiently large to stop the $\ket{A;0}$ state from evolving into the $\ket{S;0}$ state.
Figure~\ref{finite_gammas}\subref{finite_gammas(c)} 
shows the maximum
concurrence (for each time trajectory) for many different values of 
$g_1$ and $g_2$. There is a clear area of large concurrence
when 
$g_2 \approx \sqrt{3} g_1$ in accordance with the expectation from \eqref{opt_dark} in 
\secref{analytical}.  
Small
discrepancies with respect
to \eqref{opt_dark} can exist because this equation pertains to the asymptotic
concurrence and we are considering the maximum concurrence 
achieved over a finite window of time.

Note that the isolated QD population decay rates $\gamma_{p}$  
(discussed in \secref{numerical})
are
sufficiently small and are generally overwhelmed by the
Purcell decays that result from finite $\gamma_s$ and plasmon-dot
coupling factors, $g_i$. Thus, the inclusion of decay in the QD populations
has no significant effect on the results presented here for the
other parameter values considered.  
The QD dephasing terms $\gamma_{d}$, however, 
can have a more significant effect. Figure~\ref{finite_gammas}\subref{finite_gammas(d)} (similar to
Fig.~\ref{finite_gammas}\subref{finite_gammas(c)}, but with 
$\hbar \gamma_d$~=~2.0~meV) shows the 
maximum concurrence for many values of $g_1$ and $g_2$. Naturally, the maximum 
concurrence
is not as large as the 
$\hbar \gamma_d$~=~0~meV case. Furthermore, the clear peak around the
line $g_2 \approx \sqrt{3} g_1$ has been distorted, although the line still 
has some significance. Including QD dephasing effects causes the QD populations to
decay before significant entanglement can occur, unless the QDs are strongly coupled.
At small values of $g_i$, the optimal point is far from the $\sqrt{3}$ line; but
as the couplings are increased, the optimal points again fall upon the $\sqrt{3}$ line.
As mentioned above, this derivation pertains to the asymptotic values of the 
concurrence, but the dephasing does not allow the system to approach that value without
larger values of $g_i$.

\subsection{$N > 2$ Quantum Dots in the Dark} \label{ndot_dark}
We have used the analytical solution for $N$ QDs interacting
with a plasmonic system 
with no QD dephasing 
(\appref{three-state-model}) to explore how the dark
entanglement dynamics scales with increasing $N$ beyond $N=2$.
As noted in \secref{2-dots-dark}, introducing dephasing
can lead to smaller concurrences and shifts in the
optimal $g_j/g_i$ ratios, but our results should indicate
what to qualitatively expect as $N$ increases.
As in our $N = 2$ dark calculations, the initial condition
corresponds to QD1 being initially excited.

\begin{figure}
  \centering
  \includegraphics[width=.7\columnwidth]{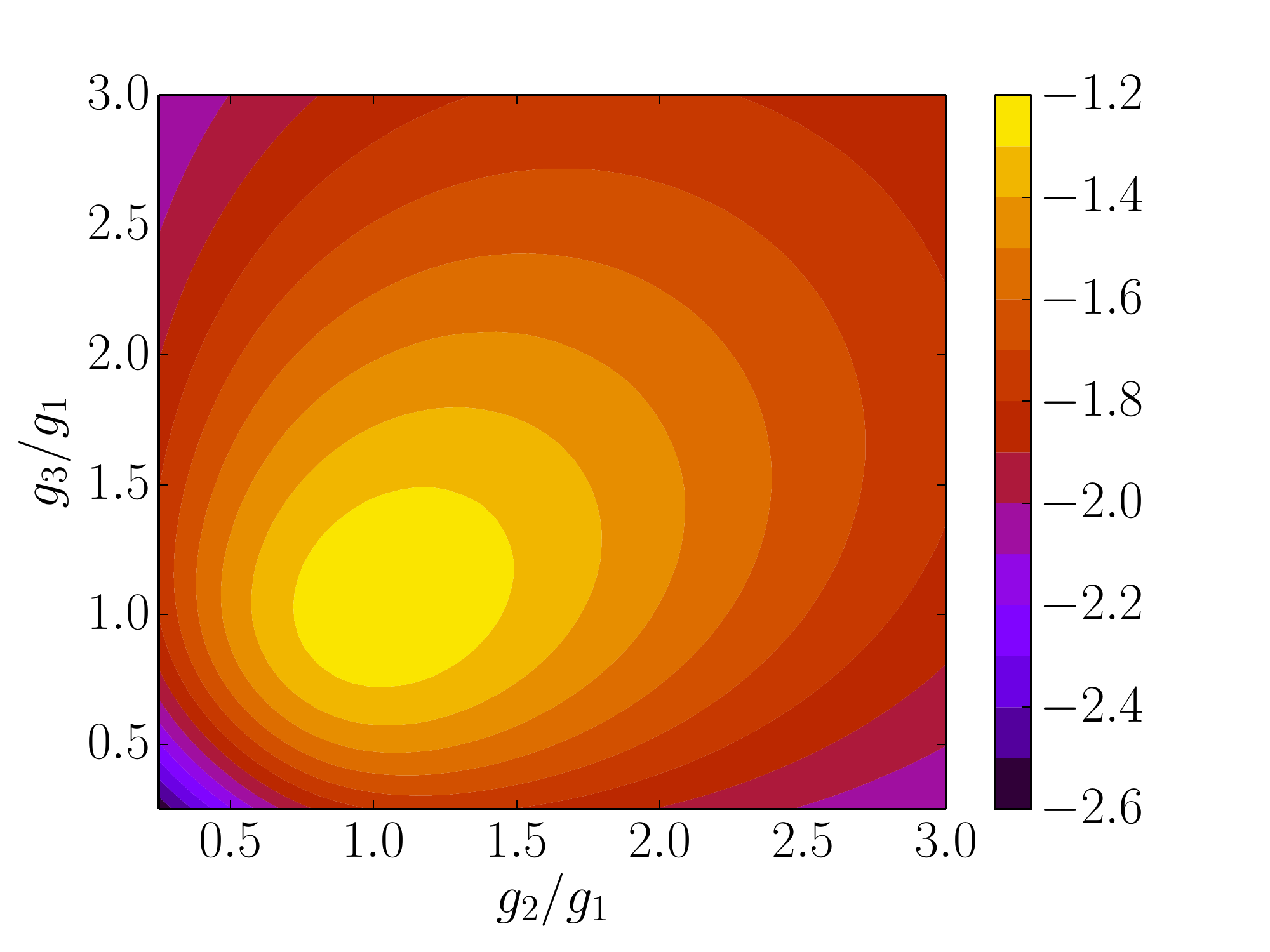}
  \caption{Asymptotic figure of merit, Eq. \eqref{eq:obj} for 
a three-QD system, with one QD initially
    excited, $\gamma_s > 0$, and $\gamma_d = 0$, as a function of the ratios of the
QD/plasmon  coupling parameters.}
  \label{3QDcontourPlot}
\end{figure}

For the $N = 3$ case, \figref{3QDcontourPlot} shows a contour map of
the asymptotic figure of merit \eqref{eq:obj} 
as a function of $g_2/g_1$ and $g_3/g_1$.
The results in this case do not depend on either $g_1$ or 
plasmon decay rate $\gamma_s$, 
provided
that the latter is positive.  
(The transient dynamics do depend on both $g_1$ and $\gamma_s$ and
can also be of interest.)
We see that the optimal concurrences
are reached at $g_2/g_1 = g_3/g_1 \approx 1.05$, which is somewhat
smaller than the $g_2/g_1 = \sqrt{3}$ ratio found for the $N = 2$ case.
The optimal values of the concurrence are $C_{1,2} = C_{1,3} \approx  
0.450$ and $C_{2,3} = 0.215$.  At 0.639, the ``direct'' concurrence
between the excited QD1 and each of the other $N-1$ dots is slightly smaller 
than the result for the two-QD system.

Although we have not derived an explicit formula, we can
evaluate the exact asymptotic dynamics of the $N$ QD case using the procedure
described in \appref{three-state-model}.  We find that for the initial
condition in question, two distinct concurrence
values always exist: a major one ($C^{\rm{maj}}$), associated with all the QD pairs that 
involve the
initially excited QD, and a smaller one ($C^{\rm{min}}$), associated with all 
the indirectly
excited pairs. Evaluation of the results for $N$ up to $N = 150$ shows that 
$C^{\rm{maj}} \approx 0.54/\sqrt{N}$,
for $N > 100$; that is, the major concurrence tends to zero, although it 
does so slowly, with an increasing number of QDs.  
In this limit, the minor concurrence decays 
somewhat
faster, with
$C^{\rm{min}} \approx 0.50/N$.
Also, the optimal concurrence figure of merit is achieved
with just one unique ratio for all the couplings, $g_{i>1}/g_1 = x$.
We find that $x \approx 1.09/\sqrt{N}$ for $N > 100$.

The optimal value of $g_2$ becomes less than $g_1$ when $N=4$, in contrast to 
the two- and three-QD 
systems, where $g_1 > g_2$. This can be explained by looking at the relative
fraction of QD pairs, $2/N$, which have $C_{i,j} = C^{\rm{maj}}$. When $N<4$, the fraction
of QD pairs that have $C_{i,j} = C^{\rm{maj}}$ is greater than $1/2$. As $N$ 
becomes larger,
more and more QD pairs have $C_{i,j} = C^{\rm{min}}$. When $g_1 > g_2$, 
there is a boost in $C^{\rm{maj}}$,
possibly at the cost of $C^{\rm{min}}$. When $N$ is large, the solution from 
minimizing the figure of merit \eqref{eq:obj} no longer favors boosting 
$C^{\rm{maj}}$; instead it increases the (more numerous) $C^{\rm{min}}$.

The state created by this mechanism, where all QDs share bipartite entanglement 
(possibly weakly) with 
all other QDs, is similar to a generalized W-state~\cite{dur-pra-2000}. 
In the W-state, all pairs of qubits have the 
same value of concurrence, and that value is as large as
possible. Thus, the W-state is the optimal state, given our figure of merit. 
According to an idea known as the monogamy of entanglement,~\cite{kim-pra-2009} 
there is an upper bound on the possible sum of 
bipartite entanglement. When $N>2$, each qubit pair can no longer be fully entangled. As
$N$ increases, the maximum bipartite concurrence for each pair in the W-state decreases as $2/N$. 
This represents a fundamental limit on the entanglement that we can achieve in our system.
The decay of the concurrence with increasing $N$ in our QD-plasmon system is 
similar to that of the W-state.
Furthermore, if we measure the initially excited QD, we can project
onto a state where QD pairs share a small degree of entanglement with each other. Taking 
a ratio of the asymptotic value of $C^{\rm{min}}$ in the projected state to the W-state concurrence
shows that each QD pair will have only $1/4$ of the concurrence of the W-state. While this
may be a small fraction, it is a constant fraction with increasing $N$, allowing
a (low-fidelity) approximate W-state to be easily created for any number of QDs. 
The addition
of decay in the QD populations
will further decrease the fidelity; but, as shown in the two-QD case,
the entanglement still persists, although at a smaller value.

\subsection{Two Quantum Dots Subjected to Ultrafast Laser Pulses}

Preparing a system in the initial state $|0,1;0\rangle$ can 
create high degrees of
entanglement, but it does not represent a simple experimental setup. 
A simpler setup is to prepare a system and excite it with a
single optical pulse. Introducing a laser pulse to a system increases the number of parameters
that the system depends on and can have a 
large effect on the value of the concurrence.~\cite{otten-prb-2015}  
% SW: I introduced this fragment because the parameters are slightly different 
% for the three QD system
For the two-QD system, 
the parameters varied include 
the laser fluence ($F$), laser duration ($\tau$), coupling strengths
($g_1$ and $g_2$), 
QD dephasing ($\gamma_d$), 
and plasmon dephasing ($\gamma_s$);
$\omega_i$, $\omega_s$, $d_i$, $d_s$, and $\gamma_p$ remain fixed.
We also constrained the parameter 
values
in a physically reasonable part of the parameter space; see \tabref{constraints}.
\begin{table}
\begin{center}
\caption{Constraints for optimization parameters.}\label{constraints}
\begin{tabular}{|l | r | r |}
  \hline
  \textbf{Parameter} & \textbf{Lower Bound} & \textbf{Upper Bound} \\ \hline 
\hline
  $\hbar g_i$ (meV) & 0 & 25 \\ \hline
  $F$ (nJ/cm$^2$) & 0 & 700 \\ \hline
  $\tau$  (fs) & 10 & 200 \\ \hline
  $\hbar \gamma_d$ (meV) & 0 & 5 \\ \hline
  $\hbar \gamma_s$ (meV) & 100 & 300 \\ \hline
\end{tabular}
\end{center}
\end{table}

We used \pounders to find optimal parameters in different parts of 
the parameter space defined in \tabref{constraints}.
% % SW: I'm not sure that this next sentence contributes anything
% We first started by optimizing a system similar to the initially excited system  
% discussed in the previous section.
We optimized the sum of the maximum value of the 
pairwise concurrence over the time horizon; other figures of merit (such as the 
sum of the integral of 
the pairwise concurrences over the time window) will be investigated in future work.

\begin{figure*}
  \centering
  \begin{subfigure}{0.5\textwidth}
    \centering
    \includegraphics[width=\columnwidth]{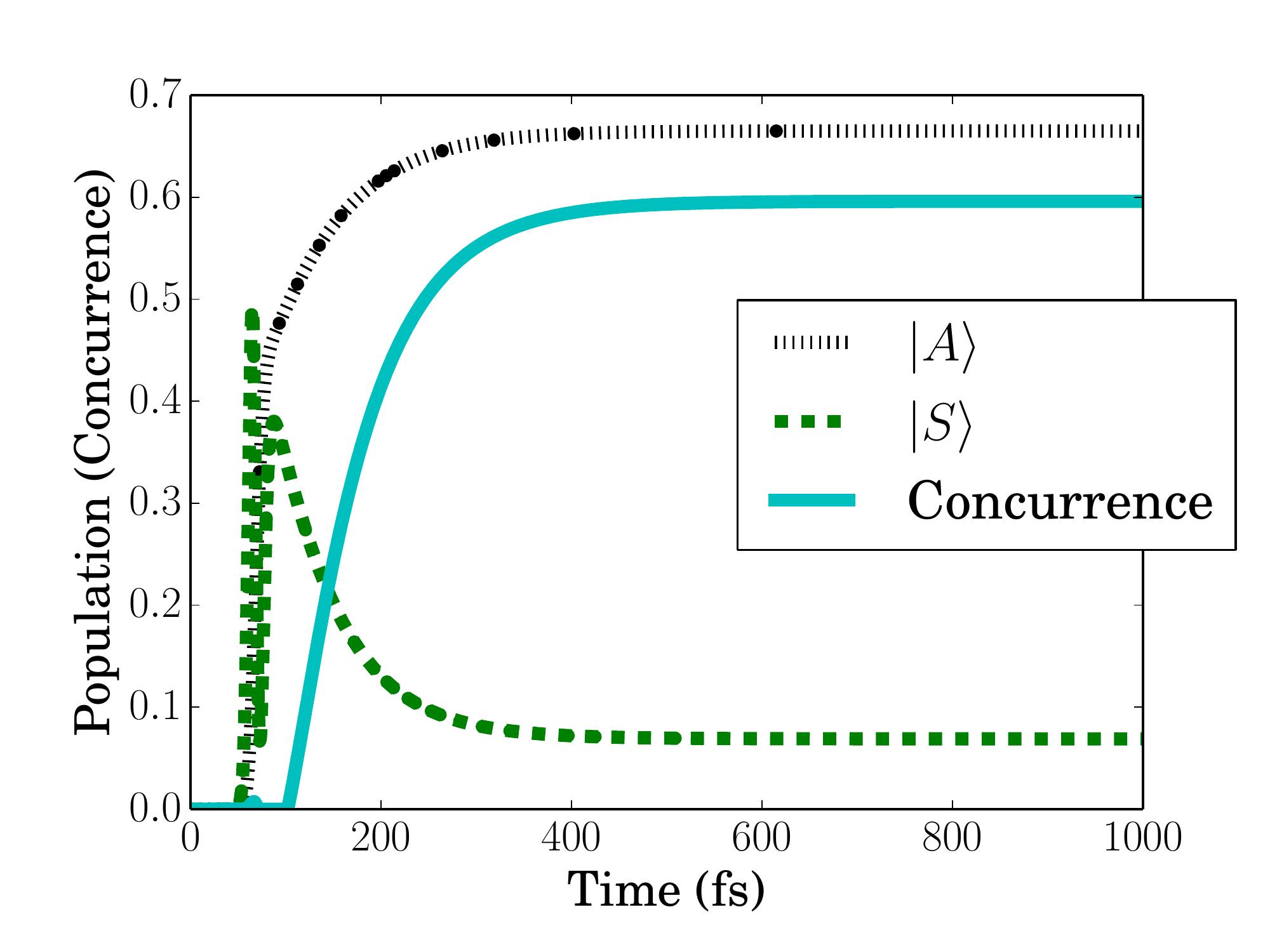}
    \caption{
}\label{2d_optimal(a)}
  \end{subfigure}%
  \begin{subfigure}{0.5\textwidth}
    \centering
    \includegraphics[width=\columnwidth]{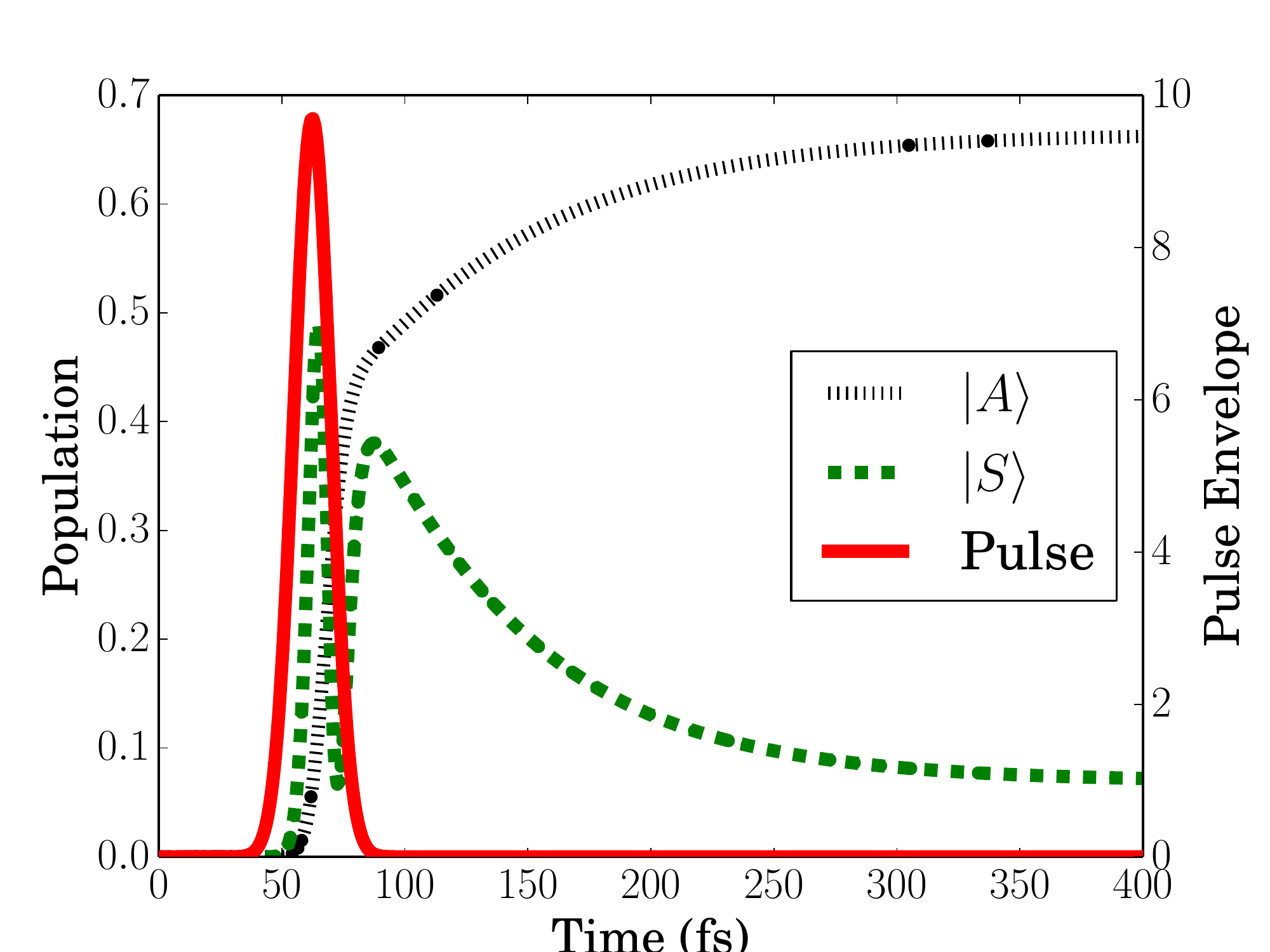}
    \caption{
}\label{2d_optimal(b)}
  \end{subfigure}%
  \\
  \begin{subfigure}{0.5\textwidth}
    \centering
    \includegraphics[width=\columnwidth]{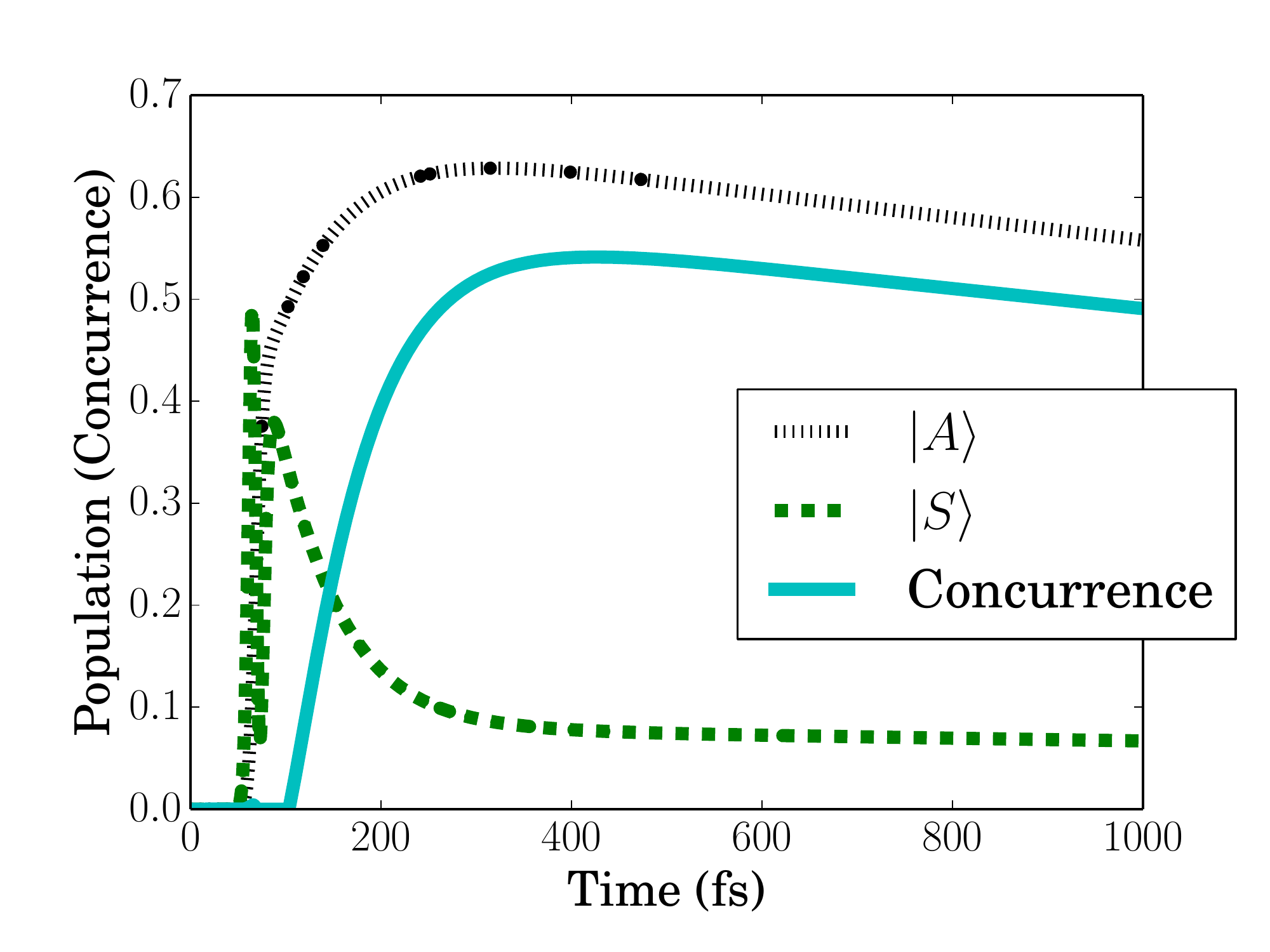}
    \caption{
}\label{2d_optimal(c)}
  \end{subfigure}%
  \begin{subfigure}{0.5\textwidth}
    \centering
    \includegraphics[width=\columnwidth]{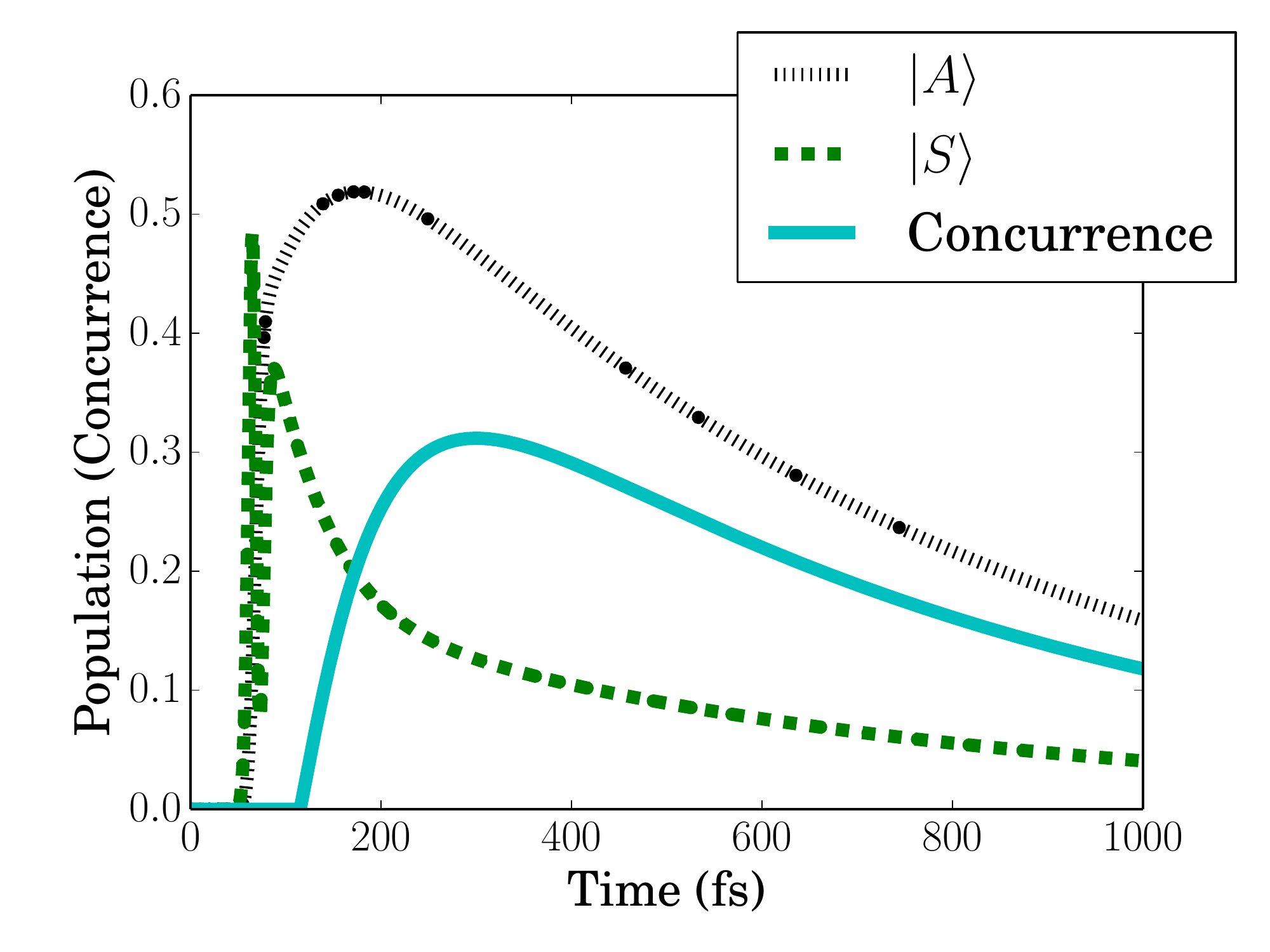}
    \caption{
}\label{2d_optimal(d)}
  \end{subfigure}%
  \caption{
    Panels (a) and (b) show the time-dependence of the 
populations of the $|A\rangle$ and $|S\rangle$ states (traced over all $|s\rangle$) 
and concurrence for  
pulsed excitation of an initially cold two-QD/surface plasmon system  with
parameters $\hbar g_1$~=~12.8~meV, $\hbar g_2$~=~24.9~meV,
$F$~=~263.4~nJ/cm$^2$, $\tau$~=~12.5~fs, $\hbar \gamma_s$~=~186~meV, 
and 
$\hbar \gamma_d$~=~0~meV. These parameters are the result of a local optimization
run.
Panels (c) and (d) keep these same parameter values except for the
QD dephasing,  which is either (c) 
$\hbar \gamma_d$ = 0.2 meV or (d) 
$\hbar \gamma_d$ = 2 meV. 
  } 
\label{2d_optimal}
\end{figure*}

%% {\bf SKG remark: For Fig. 4 and discussion: First, the optimal parameters should be given in caption
%% and  not in text -- that would be consistent with what SW remarked on
%% and the text would flow better.
%% However, I'm confused because are there not different optimal parameters for
%% different panels in Fig. 4?  
%% The text discussion seems to be implying that at least the
%% panel (d) has g1 $<$ g2 but text also states g1 $>$ g2 for (a).  So some
%% clarification and putting in someplace the optimal values is needed, along
%% with connecting with the relative g's and the boost arguments of previous
%% sections.} 
%% occurs when $\hbar g_1$~=~24.9~meV, $\hbar g_2$~=~12.8~meV,
%% $F$~=~263.4~nJ/cm$^2$, $\tau$~=~12.5~fs, $\hbar \gamma_s$~=~186~meV,
%% and $\hbar \gamma_d$~=~0~meV. 
The evolution of the pairwise concurrence and the states' 
populations for a locally 
optimal result are given 
in Fig.~\ref{2d_optimal}\subref{2d_optimal(a)} and are seen to behave similarly to the  
dark case with one initially excited QD shown in  
Fig.~\ref{finite_gammas}\subref{finite_gammas(a)}. 
In contrast to that system, the plasmon population (not shown) 
reaches a much higher value of nearly 10 in this system. 
(The $\ket{A}$ and $\ket{S}$  state populations shown in 
Fig. ~\ref{2d_optimal} result
from tracing the density matrix over all plasmon quantum numbers.)
We previously discovered that we had to allow $g_1$ and
$g_2$ to differ in order to create large amounts of concurrence because doing so allowed the system
to approximate the $|0,1\rangle$ state, creating a highly entangled state, with the 
proper parameter choices.~\cite{otten-prb-2015} We noted there
that the less-strongly coupled QD achieved a higher population after 
the pulse concluded. 
The boost in the $\ket{A;s=0}$ population 
we describe in this paper for the case 
$g_1 < g_2$ (see discussion of \figref{cyclic_evolution})  
is also present, 
helping raise
the concurrence higher and thereby allowing the pulsed
case to reach levels of concurrence similar to those for the dark case. 
This is clearly seen in Fig.~\ref{2d_optimal}\subref{2d_optimal(b)},
where the initial time scale has been expanded and the pulse
envelope is also displayed. After the pulse ends, the 
$\ket{S}$ state
begins to decline, but the $\ket{A}$ state grows, because of their indirect coupling.
This boost of the $\ket{A}$ state is the same as seen in the dark case.
Figures~\ref{2d_optimal}\subref{2d_optimal(c)}--\ref{2d_optimal}\subref{2d_optimal(d)}
also show this same parameter set with 
larger values of $\gamma_d$. The maximum value of the pairwise
concurrence strongly depends on the QD dephasing, $\gamma_d$. This 
dependence is not surprising, because longer 
coherence times are almost always associated with larger degrees of (and 
longer-lived) 
entanglement. Figure~\ref{2d_optimal}\subref{2d_optimal(c)} shows the system at 
$\hbar \gamma_d$~=~0.2~meV (approximately liquid helium temperatures), while 
Fig.~\ref{2d_optimal}\subref{2d_optimal(d)} shows the system at 
$\hbar \gamma_d$~=~2.0~meV (approximately liquid nitrogen
temperatures). The loss in concurrence from 
$\hbar \gamma_d =$ 0~meV to $\hbar \gamma_d =$ 0.2~meV
is only about 10$\%$,
but it is almost 50$\%$ when 
$\hbar \gamma_d$ is raised to 2.0~meV. Generally, the concurrence increases
with decreasing $\gamma_d$.

%% We can calculate the optimal incoming field strength,
%% \begin{equation}\label{opt_E0}
%%   E_0 = \frac{n\pi \hbar \gamma_s}{\sqrt{2} g \mu_s \tau},
%% \end{equation}

\begin{figure}
  \centering
  \includegraphics[width=\columnwidth]{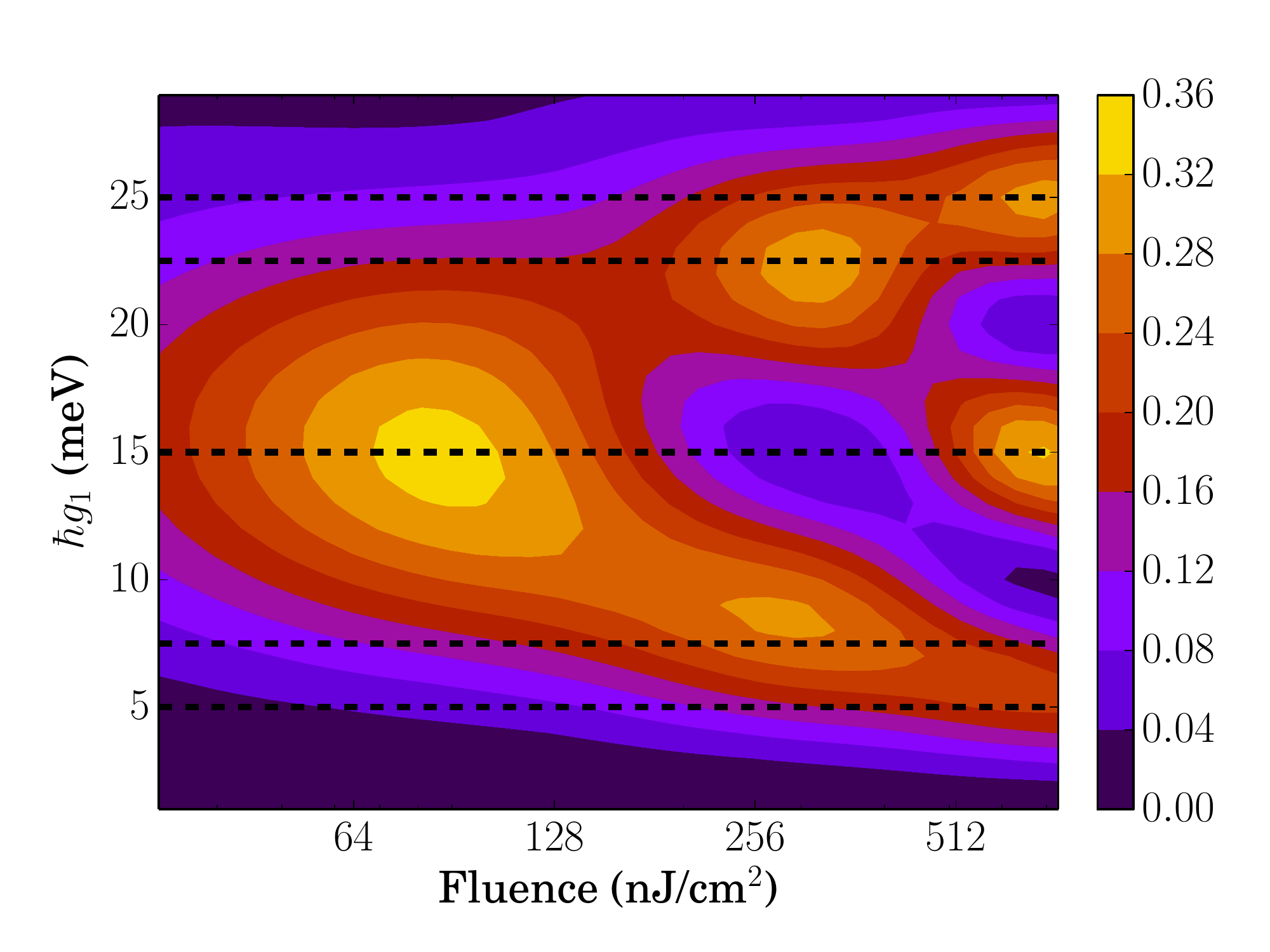}
  \caption{Maximum concurrence for a parameter sweep of the two-QD system, with $\hbar g_2$~=~30~meV,
    $\tau$~=~20~fs, 
$\hbar \gamma_s$~=~150~meV, 
$\hbar \gamma_d$~=~2~meV. The dashed lines represent coupling ratios 
obeying Eq. \eqref{opt_pulse}.}
  \label{contourPlot}
\end{figure}

\figureref{contourPlot} shows the maximum concurrence over our time window as
$g_1$ and $F$ vary, for fixed pulse duration ($\tau$) and coupling strength of
the second QD ($g_2$). An interesting consequence of the
Rabi oscillations is bifurcations of the areas of high concurrence. At small laser fluences, 
given $g_1 < g_2$, there is only
one region of high concurrence
corresponding to one QD undergoing a half
Rabi oscillation and the other undergoing one oscillation; that is,
the $m = n = 1$ case from \secref{analytical} that was
predicted to maximize entanglement.
As the laser fluence is increased, the region of high concurrence splits into two regions, as the
more-strongly coupled QD approaches two full Rabi oscillations. The 
less-strongly coupled QD can now either go through half or three-halves 
Rabi oscillations to end
up in the excited state. This region bifurcates again, as the second dot 
approaches three Rabi oscillations.
This analysis works for the two-QD and three-QD systems we present in
this paper, but it gives a relationship only between two of the parameters, $g_1$ and $g_2$. 
Since we have
many other parameters to optimize over, \pounders is used to find local optima of the
maximum concurrence.

\subsection{Three Quantum Dots Subjected to Ultrafast Laser Pulses}

Since the QDs are assumed to be coupled to the plasmon but not to each other, 
adding a QD increases the number of parameters only by one ($g_3$, the
new QD's coupling to the plasmon). More importantly, the size of the Hilbert
space needed for the simulation increases by a factor of 2, more than quadrupling the
simulation's run time and making the
optimization algorithm's ability to quickly find locally optimal solutions 
even more important.
Here we present two locally optimal points for a three-QD system. The 
QD dephasing, $\hbar \gamma_d$, is fixed to 0.2~meV, since this approximates a physically realizable
system at liquid helium temperatures. 

\begin{figure*}[!ht]
  \centering
  \begin{subfigure}{0.5\textwidth}
    \centering
    \includegraphics[width=\columnwidth]{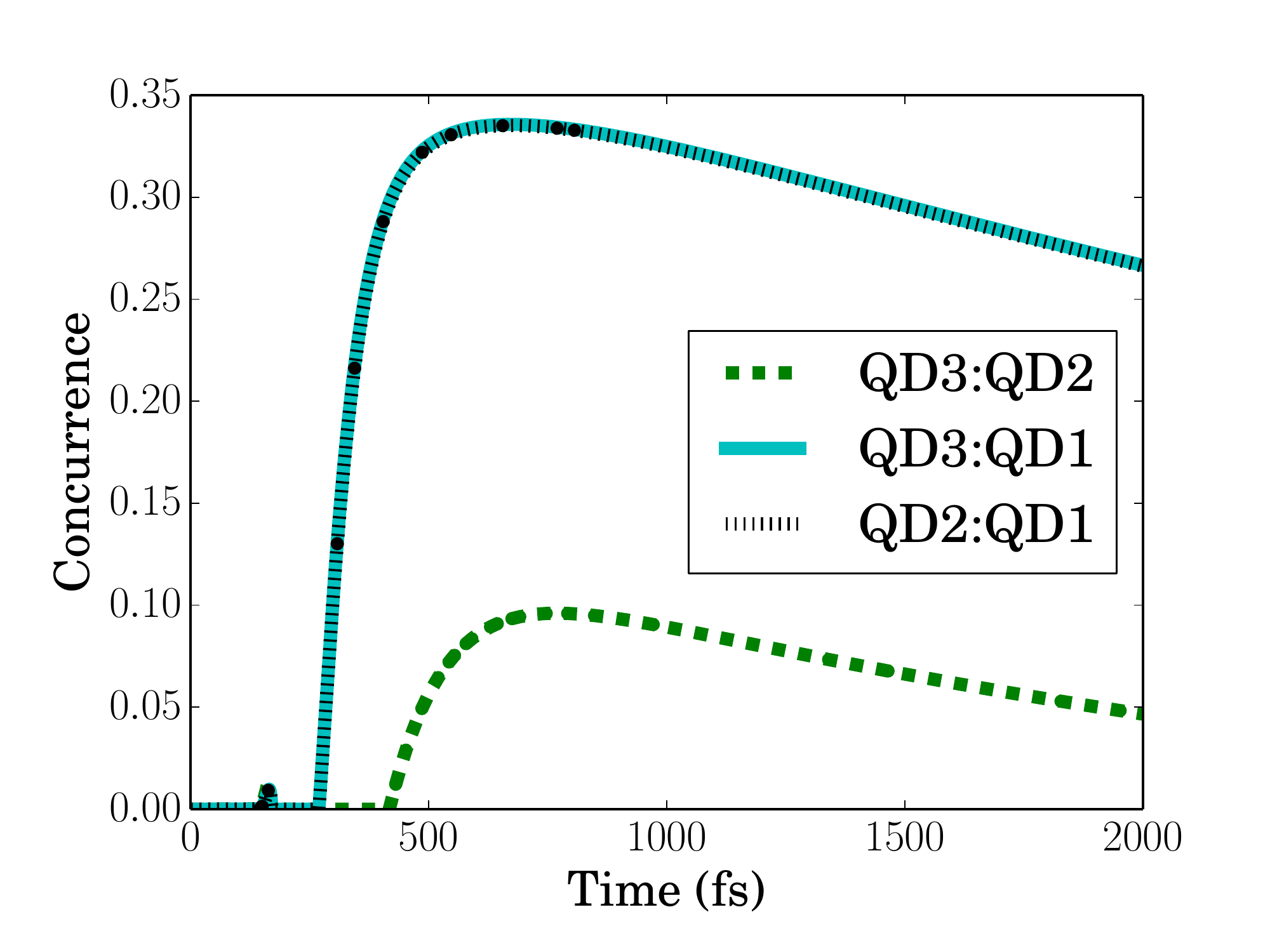}
    \caption{}\label{3qd_param4(a)}
  \end{subfigure}%
  \begin{subfigure}{0.5\textwidth}
    \centering
    \includegraphics[width=\columnwidth]{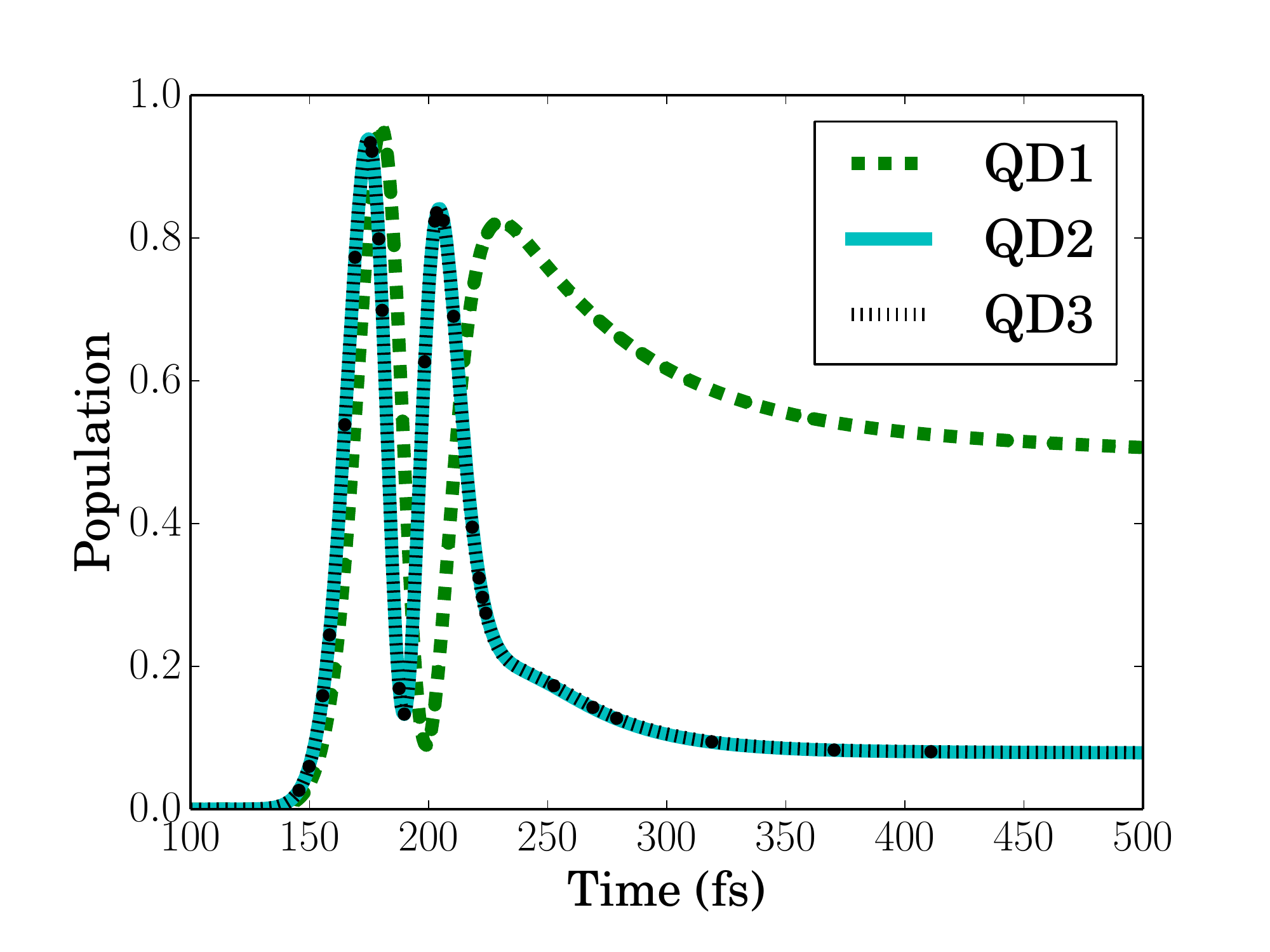}
    \caption{}\label{3qd_param4(b)}
  \end{subfigure}%
  \\
  \begin{subfigure}{0.5\textwidth}
    \centering
    \includegraphics[width=\columnwidth]{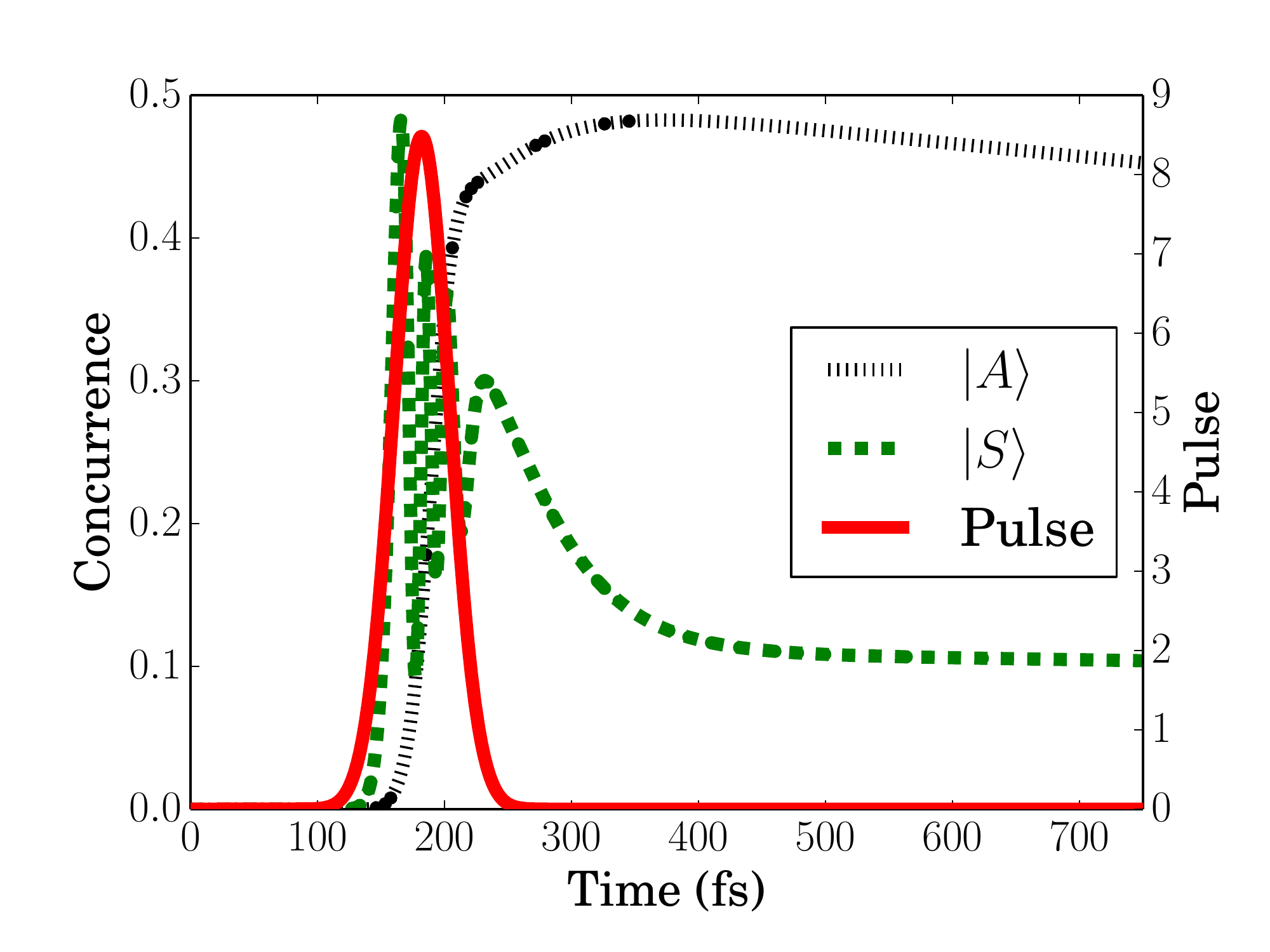}
    \caption{ 
}\label{3qd_param4(c)}
  \end{subfigure}%

  \caption{
    Populations and concurrences for the final parameters from a local optimization 
    run on the three-QD system with $\hbar \gamma_d$ fixed at 0.2~meV.
    The final parameters 
    are $\hbar g_1$~=~14.6~meV, 
$\hbar g_2$~=~19.3~meV, $\hbar g_3$~=~19.3~meV, 
$F$~=~587.0~nJ/cm$^2$, $\tau$~=~36.4~fs, and
    $\hbar \gamma_s$~=~180.4~meV (with $\hbar \gamma_d$ fixed at 0.2~meV). 
Panel (a) shows the various bipartite concurrences and panel (b) shows
the QD excitation probabilities.  Because $g_2$ = $g_3$,
the QD3:QD1 and QD2:QD1 concurrences are identical, as are the
QD2 and QD3 excitation probabilities. Panel (c) shows the 
time-dependent probabilities of the
 $\ket{S}$ and $\ket{A}$ states associated with either the 
QD3:QD1 or QD2:QD1 subsystems and the pulse envelope.}
\label{3qd_param4}
\end{figure*}

\figureref{3qd_param4} shows the populations of the QDs and their
pairwise concurrences
for the system parameters returned from 
a local optimization run. 
This system is analogous to the
two-QD systems discussed above, since $g_2=g_3$. QD2 and QD3 undergo two 
Rabi flops, and QD1
undergoes one-and-a-half Rabi flops.
Accordingly, $g_2/g_1 = g_3/g_1 = 1.322 \approx \frac{4}{3}$, 
as predicted by \eqref{opt_pulse}. The boost of the population of the 
$\ket{A}$ state is also apparent in this
system. Shortly after the pulse has concluded, the $\ket{A}$ state is still rising, while the
$\ket{S}$ state decays. The boost of the $\ket{A}$ state eventually finishes 
and the $\ket{S}$ and $\ket{A}$ states
then decay at similar rates.
Aside from having a much larger concurrence than presented previously, 
the pulsed three QD simulations presented in this paper are 
also interesting because their coupling parameters are smaller
and represent a more physically reasonable system than do our previous results.~\cite{otten-prb-2015}

\begin{figure*}[!ht]
  \centering
  \begin{subfigure}{0.5\textwidth}
    \centering
    \includegraphics[width=\columnwidth]{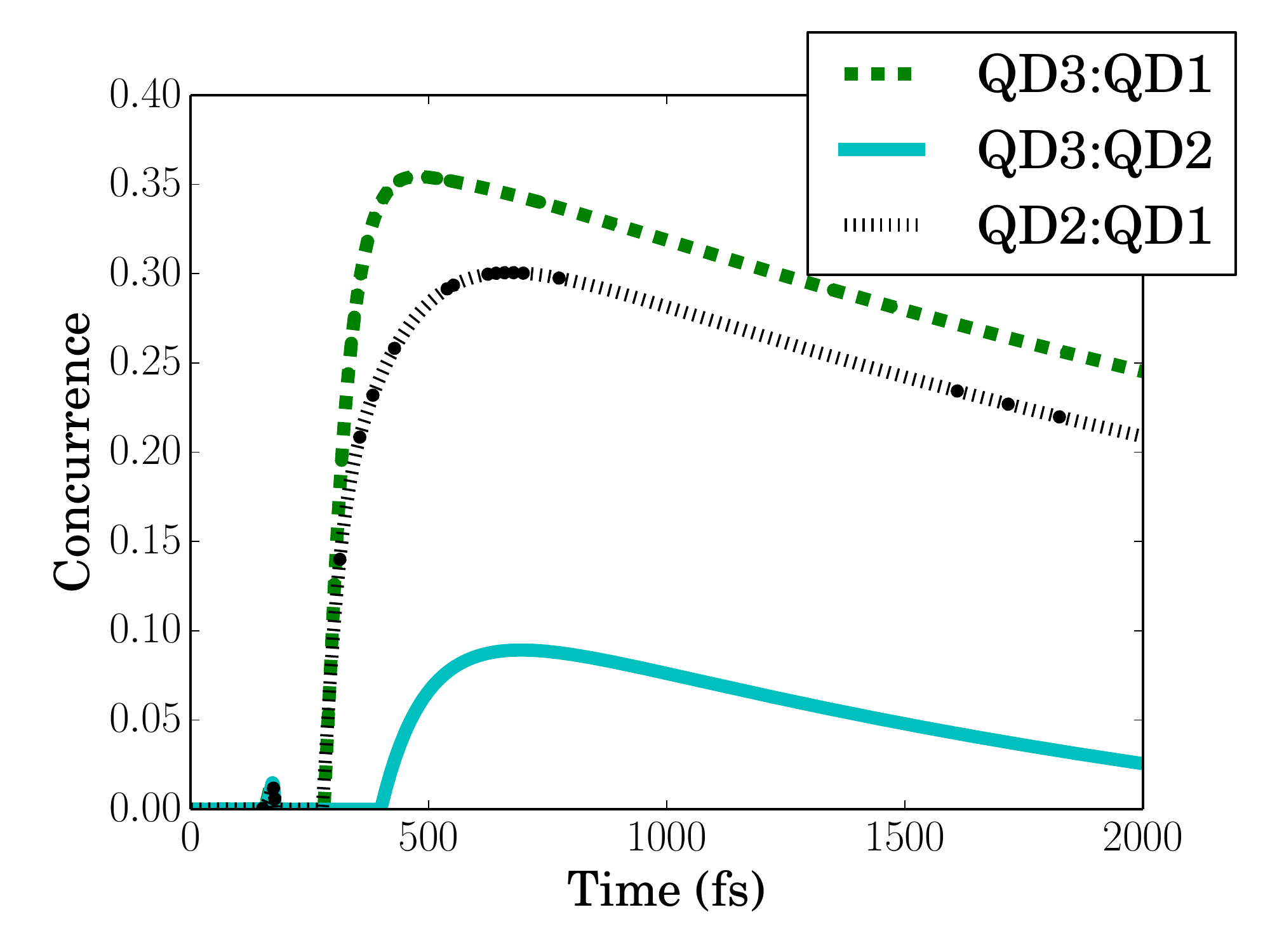}
    \caption{}\label{3qd_param5(a)}
  \end{subfigure}%
  \begin{subfigure}{0.5\textwidth}
    \centering
    \includegraphics[width=\columnwidth]{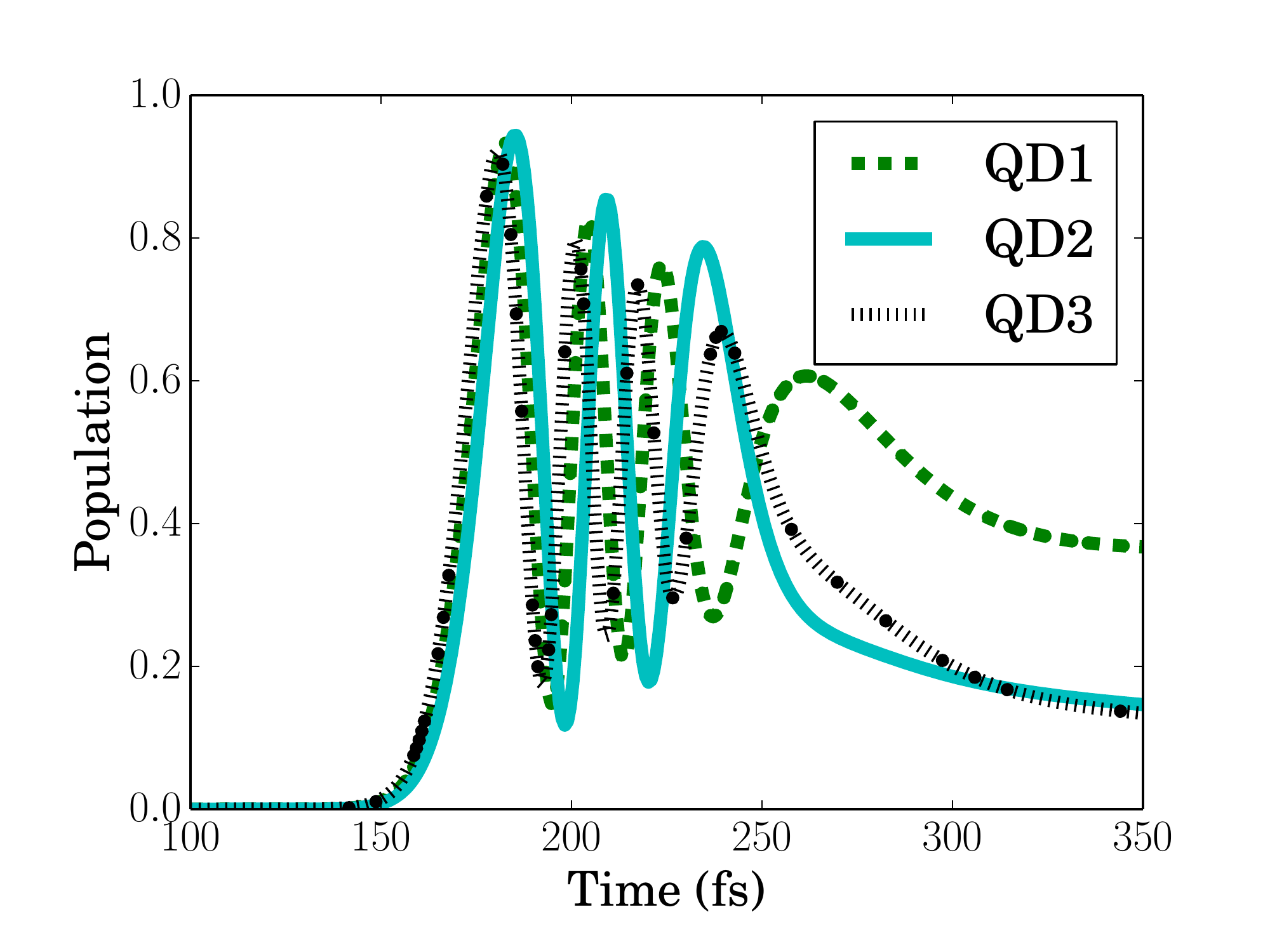}
    \caption{}\label{3qd_param5(b)}
  \end{subfigure}%
  \\
  \begin{subfigure}{0.5\textwidth}
    \centering
    \includegraphics[width=\columnwidth]{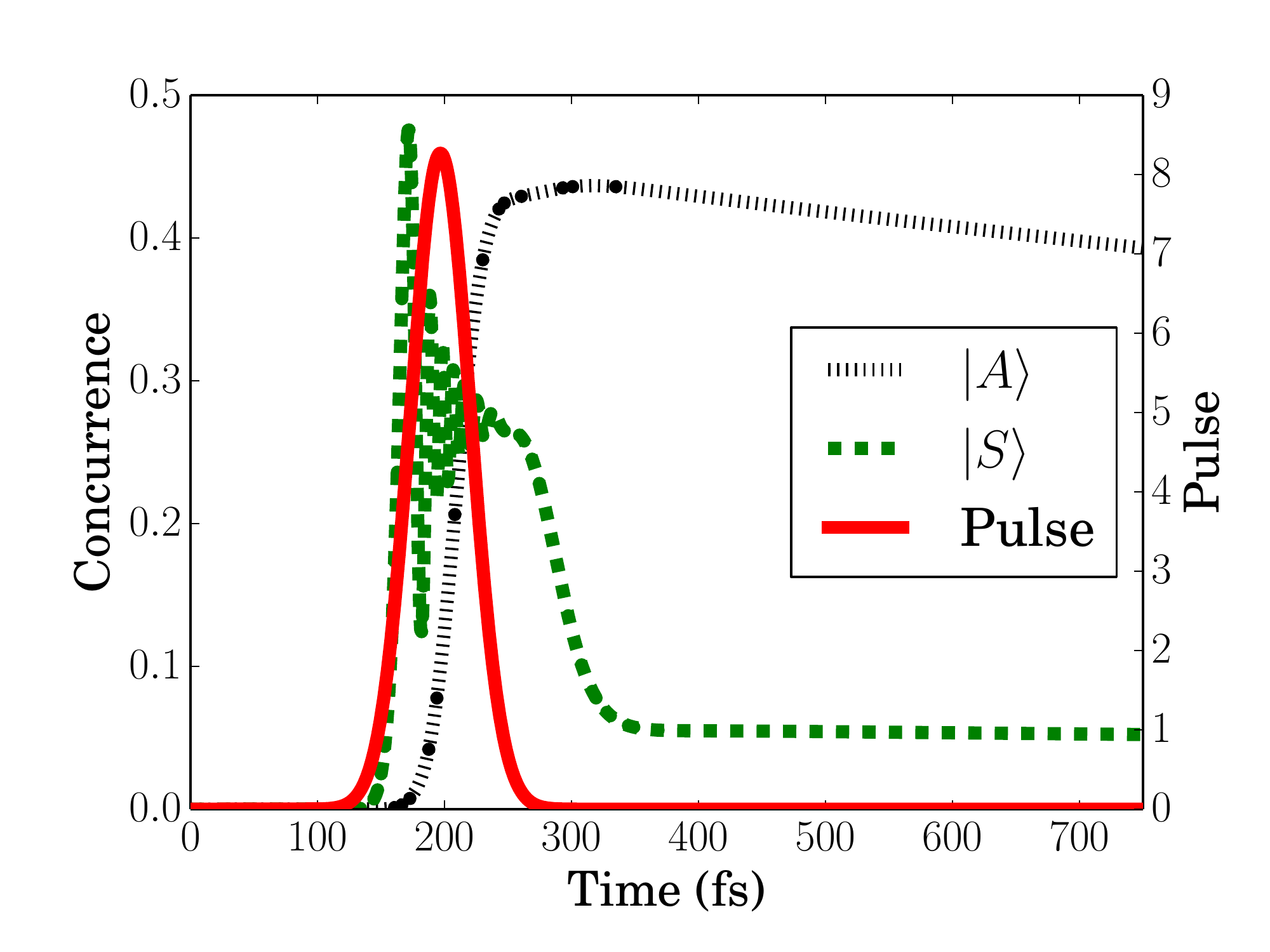}
    \caption{}\label{3qd_param5(c)}
  \end{subfigure}%
 \begin{subfigure}{0.5\textwidth}
    \centering
    \includegraphics[width=\columnwidth]{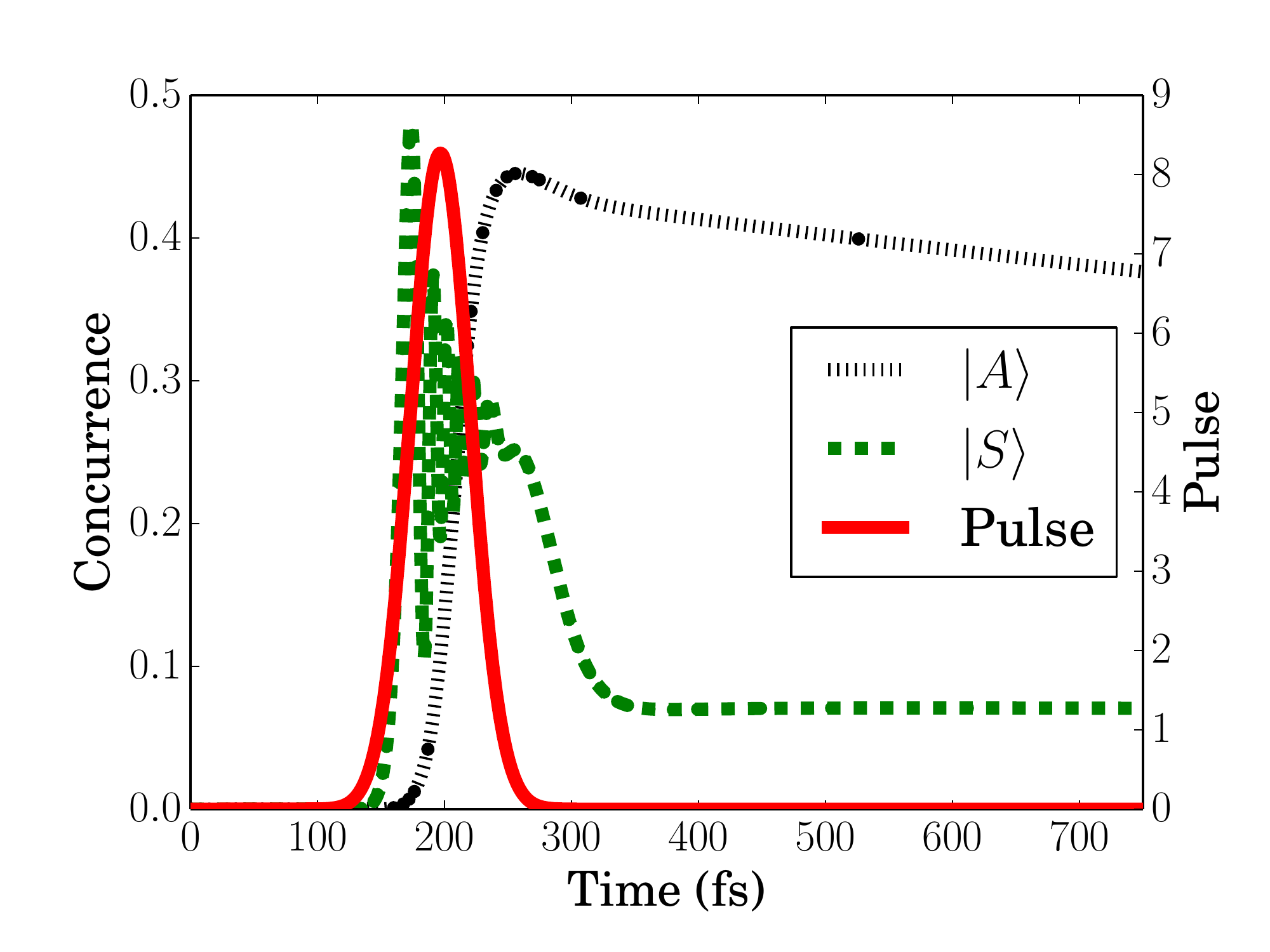}
    \caption{}\label{3qd_param5(d)}
  \end{subfigure}%

  \caption{
    Populations and concurrences for the final parameters from a second local optimization 
    run on the three-QD system with $\hbar \gamma_d$ fixed at 0.2~meV.
    The final parameters are
    $\hbar g_1$~=~19.0~meV, $\hbar g_2$~=~16.3~meV,  $\hbar g_3$~=~21.7~meV, 
    $F$~=~603.4~nJ/cm$^2$, $\tau$~=~39.4~fs, and
    $\hbar \gamma_s$~=~107.7~meV (with $\hbar \gamma_d$ fixed at 0.2~meV).
Panel (a) shows the various bipartite concurrences and panel (b) shows the QD
excitation probabilities.
The pulse envelope and the  populations of the $\ket{S}$ and $\ket{A}$ 
states are shown for the QD3:QD1 pair in panel (c) and the QD2:QD1 pair
in panel (d). 
  } 
\label{3qd_param5}
\end{figure*}

\figureref{3qd_param5} shows the populations and concurrences for a system with 
the best parameters from a different local optimization
run. 
The QDs in this system all have different coupling values (as opposed
to the previous example where $g_2 = g_3$), leading to three different pairwise
concurrences, even though the populations of QD2 and QD3 are similar in
value.  QD3 undergoes four Rabi flops, while QD1 undergoes 
three-and-one-half
Rabi flops, leading to $g_3/g_1 = 1.142 \approx \frac{8}{7}$. Additionally, QD2
undergoes three Rabi flops, leading to $g_2/g_1 = 0.858 \approx \frac{6}{7}$.
Both these pairs agree with \eqref{opt_pulse}.
The QD3:QD1 subsystem, 
where the
excited QD has a smaller coupling value, 
exhibits the boost 
of the population of the $\ket{A}$ state described previously,
as shown in Fig.~\ref{3qd_param5}\subref{3qd_param5(c)}. Contrast this
with the QD2:QD1 subsystem, Fig.~\ref{3qd_param5}\subref{3qd_param5(d)}, which does not experience the boost,
since the excited QD has a larger coupling. After the pulse concludes, the 
$\ket{S}$ state undergoes a similar evolution, but the $\ket{A}$ 
state is different. In the
QD3:QD1 subsystem the $\ket{A}$ state increases after the pulse has concluded, but in the
QD2:QD1 subsystem the $\ket{A}$ state only decreases. As a result, the QD3:QD1 
concurrence ends up 
greater than the QD2:QD1 state, even though the populations of QD2 and QD3 end up at similar
values.

We note that the Rabi-flop mechanism of Eq. \eqref{opt_pulse} singles out one
QD (the one undergoing $m-\frac{1}{2}$ Rabi flops) to become strongly entangled with all other
QDs, while the other two QDs become strongly entangled only with the excited QD and
become weakly entangled with each other. This is
an inherent limit of the prescription described. Remarkably,
entanglement between the two approximate ground state QDs reaches the level that it does,
even for identical QDs (i.e., a concurrence of 0.1 in \figref{3qd_param4}), but the entanglement is still
much smaller than the entanglement they share with the excited QD.  Since the pulse 
approximately prepares the 
same state studied in \secref{ndot_dark} and the Rabi-flop mechanism can be used for 
$N$ quantum dots, we can again project onto a (low-fidelity) approximate W-state,
where all pairs of QDs share the same amount of bipartite entanglement. This approach
is more experimentally feasible than having a previously excited QD, but 
the bipartite concurrence values will be lower than those of \secref{ndot_dark}.

\begin{table}
\begin{center}
\caption{Summary of optimization run results
for a laser pulse interacting with a system composed of
two and three quantum dots.
An``$^f$'' denotes that the parameter 
value was fixed and therefore not optimized over.}\label{tab:summary}
\begin{tabular}{|l | r | r | r |}
  \hline
  \textbf{} & \textbf{Two QDs} & \textbf{Three QDs, 
Solution 1} & \textbf{Three QDs, Solution 2} 
\\ \hline 
\hline
 $\hbar g_1$ (meV)  	&  12.8	& 14.6 & 19.0 \\
 $\hbar g_2$ (meV)  	&  24.9	& 19.3 & 16.3 \\ 
 $\hbar g_3$ (meV)  	&  ---	& 19.3 & 21.7 \\ 
 $F$ (nJ/cm$^2$) & 263.4& 587.0 & 603.4 \\ 
 $\tau$  (fs) 	&  12.5	& 36.4 & 39.4 \\ 
 $\hbar \gamma_d$ (meV) &  0	& 0.2$^f$ & 0.2$^f$  \\ 
 $\hbar \gamma_s$ (meV) & 186	& 180 & 108 \\ \hline
   Figure & \figref{2d_optimal} & \figref{3qd_param4} & \figref{3qd_param5} \\ 
\hline
   
  Maximum bipartite concurrence & 0.60 & 0.34 & 0.35 \\ \hline
\end{tabular}
\end{center}
\end{table}

\section{Concluding Remarks}
We provide a detailed explanation of the origins and optimization of
bipartite (or pairwise) 
entanglement in two,
three, and an arbitrary number of QDs coupled 
to a plasmonic system. We analyze systems with an initially excited state as well as initially
unexcited systems excited by a laser pulse. We vary the QD-plasmon coupling 
values 
(which represent a QD's distance from the plasmonic system), as 
well the
femtosecond pulse parameters and dephasing rates to explore entanglement generation.
By utilizing the full density matrix master equation, we are able to study the 
entanglement (via concurrence) of many different systems.

In the case of two QDs, two mechanisms are identified as the source 
of the entanglement generation: the differing decay rates of the $\ket{S}$ and
$\ket{A}$ states (previously identified)~\cite{otten-prb-2015} and a new 
mechanism 
involving an indirect coupling between the $\ket{S}$ and $\ket{A}$ states that
leads to a boost 
in the $\ket{A}$ state population. 
With no dephasing or decay, high 
degrees of entanglement can be generated by having near-unity populations of
either the $\ket{A}$ or $\ket{S}$ states. When plasmon decay is added, however, 
the entanglement generated from
$\ket{A}$ is much higher than that of $\ket{S}$.
A simple analysis including plasmon decay but neglecting
QD dephasing predicts that the asymptotic concurrence
is maximized when $g_2/g_1 = \sqrt{3}$ in the dark case; 
calculations show that this relation is still useful when 
QD dephasing is considered.

The two entanglement-generating mechanisms are most apparent when the system is initially prepared
with equal amounts of $\ket{S;0}$ and $\ket{A;0}$, which is most easily achieved by having
one excited QD and the other QD in the ground state. 
This dark case may be contrasted with attempts to generate 
entanglement in initially unexcited systems by using laser pulses.
On the basis of optimization, we find that 
only certain sets of parameters generate an analog to the dark 
case. 
In particular, certain rational values of 
the ratio of the QD-plasmon couplings, $g_i/g_j$, lead to results comparable to the dark case.
These ratios can be understood by analyzing the underlying Rabi flops of the component
QDs; the target final state, after the pulse, consists of one QD excited and
the other QD in the ground state. To achieve this, one QD undergoes $m - \frac{1}{2}$
Rabi oscillations, leaving it in an approximate excited state, while the other QD undergoes
$n$ full oscillations, leaving it in an approximate ground state. This method will work
for any pair of QDs, even if that pair is part of a larger system of $N$ QDs.

In the case of three QDs, we optimize the sum of the bipartite 
concurrences
among all of the pairs. Several local maxima corresponding to different sets of system parameters
are obtained, and we present two in this paper. One 
(``solution 1'')
was analogous to the two-QD 
systems discussed above, with $g_1 < g_2 = g_3$, while the 
other (``solution 2'') 
had $g_2 < g_1 < g_3$; both exhibited the entanglement generation
mechanisms described above.
The parameters for all three optimal systems are listed in \tabref{tab:summary}. 
The ratios of the $g_i$ values in the highly entangled pairs of these three-QD systems 
follow the simple rules derived from the Rabi-flop analysis.

We also extended our results to $N$ QDs, with some simplifying assumptions, such as no
QD population decay and a single initially excited QD. For any number of QDs,
all pairs of QDs will become entangled. 
However, since this mechanism relies on one QD being in the excited state
and the rest of the QDs being in the ground state, these mechanisms
can strongly entangle only a fraction ($2/N$) of the
pairs of QDs. Using the simple rules laid out in this paper for a large number of QDs
results in the excited QD being strongly entangled with
all other QDs, but all the ground state QDs will be strongly entangled only with
the excited QD and only weakly entangled with each other. Since all QDs share some amount
of bipartite entanglement with all other QDs, the resulting state is similar
to a generalized W-state and, with a measurement of the excited QD, can be
projected to a state where all pairs of QDs share the same amount of bipartite concurrence 
(though this projected state only has $1/4$ of the bipartite concurrence of 
a true W-state).  

Additionally, this procedure could generate certain types 
of cluster states. A cluster state is a graph in which qubits are 
represented by nodes on a graph, and an edge between two nodes represents
entanglement between the two qubits.~\cite{briegel-prl-2001} The W-state
would be a cluster state represented by a complete, fully connected graph. 
An important subclass of cluster states is the star state, where a central
node is connected to all other nodes (or, a single qubit is entangled to 
all other qubits but the other qubits are not entangled with each other).
A four-qubit star cluster can be used for universal quantum computing.~\cite{bell-njp-2013}
In our model, we have an approximate star cluster for $N$ QDs, since the initially excited
QD is strongly entangled with all other QDs. Although we showed $N$ QD 
results only for  
a specific initial starting condition, we also show how this state
can be prepared for $N$ QDs from a single, ultrafast laser pulse. The rules
for the ratios of the coupling strengths based on the number of Rabi flops
can be used to define an appropriate set of parameters to prepare an 
approximate form of the specific initial starting condition studied.

Further studies of such systems that better approximate the W-state
are planned, as are studies of the
entanglement between all qubits of the system (rather than just pairs), which would be
similar to the GHZ-state,~\cite{dur-pra-2000} which represents entanglement where
all of the qubits are mutually entangled with each other (rather than just sharing
bipartite entanglement with other qubits). The W- and GHZ-states represent two 
mutually exclusive examples of multipartite entanglements and allow entanglement
to be used as a quantum information resource in different ways.~\cite{dur-pra-2000}
We will also apply the same ideas and methods to other systems,
such as nitrogen vacancies in diamond and superconducting qubits.

\begin{acknowledgments}
This work was performed at the Center for Nanoscale Materials, 
a U.S.\ Department of Energy, Office of Science, Office of Basic Energy 
Sciences User Facility 
and supported by the 
U.S.\ Department of Energy, Office of Science, Office of Advanced Scientific 
Computing Research 
(both under Contract No.~DE-AC02-06CH11357).
We thank Ron Shepard for helpful suggestions concerning some of the
analytical analysis presented in this paper.
We thank
Todd Pittman and Jason Kestner for helpful discussions.
\end{acknowledgments}

\appendix
\section{Three and $N+1$ State Models}
\label{three-state-model}
The ``dark'' problem, namely, to determine the 
dynamics  of $N$ QDs and a plasmonic system 
that results 
from a given initial condition without any applied laser pulse,
can be solved analytically if the
initial condition is not too energetic and QD dephasing is neglected. 
An example of such a system would be if there is just one
quantum of excitation within the QD manifold and a cold plasmonic system. 
The analytical solution  
is made possible because 
under such conditions
a time-dependent Schr\"{o}dinger equation involving an
effective, non-Hermitian Hamiltonian can be employed and the latter
can be represented by an $(N+1) \times (N+1)$ matrix with a simple structure.
First we illustrate such a solution in detail for the case of $N = 2$.
We then present the general $N + 1$ state solution.

For two QDs interacting with a plasmonic system,  we wish to solve for
the time evolution of $\ket{\Psi (t)}$ satisfying  
\begin{equation} \label{tdse}
i \hbar \frac{\partial}{\partial t} \ket{\Psi (t)} = \hat{H} \ket{\Psi (t)},
\end{equation}
where
\begin{equation}
  \begin{array}{rl}
  \ket{\Psi (t)} =& c_0(t) \ket{q_2=0,q_1=0;s=1} \\
                &+ \;c_S(t) \ket{S;s=0} + c_A(t) \ket{A;s=0}.
  \end{array}
\end{equation}
We refer to the three states 
$\ket{q_2=0,q_1=0;s=1}$,
                $\ket{S;s=0}$, and  $\ket{A;s=0}$ as the zero-order
basis.
This limited basis is adequate for describing an initial condition
that involves any superposition of these three states, such
as the case of one QD being excited and the plasmonic system and other
QD being cold.
%For brevity, we will sometimes use the condensed notation for the 
%three zero-order
%basis states $\ket{0} = \ket{q_2=0,q_1=0;s=1}$, $\ket{S} = \ket{S;s=0}$, and
%$\ket{A} = \ket{A;s=0}$. 
With the definitions of the basis states in the text, \eqref{primitive}, 
\eqref{sym_state}, and \eqref{asym_state}, and the Hamiltonian operator,
\eqref{hamiltonian}, the corresponding $3 \times 3$ Hamiltonian matrix 
of the zero-order basis representation is 
\begin{equation} \label{hambase}
\boldsymbol{H} = \hbar
  \begin{bmatrix}
    \omega_0 & \alpha & \beta  \\
    \alpha & \omega_0 - i \epsilon & 0 \\
    \beta  &  0    & \omega_0
  \end{bmatrix}.
\end{equation}
The QD and plasmon transition
frequencies are assumed to be equal, $\omega_1 = \omega_2 = \omega_s$,  
the coupling between $\ket{0,0;1}$ and  
$\ket{S;0}$ is 
\begin{equation}
\alpha = \frac{1}{\sqrt{2}} \big(  g_1 + g_2 \big),
\end{equation}
and the coupling between
$\ket{0,0;1}$ and  
$\ket{A;0}$ is 
\begin{equation}
\beta = \frac{1}{\sqrt{2}} \big(  g_1 - g_2 \big).
\end{equation}
We assume no direct coupling between $\ket{S;0}$ and $\ket{A;0}$.
Notice that in \eqref{hambase}, we have added an imaginary
part $-i \epsilon$ to the diagonal matrix element associated
with $\ket{0,0;1}$.  With $\epsilon = \gamma_s/2$ this term
represents the dissipative loss of the plasmonic system. 

Introducing the more slowly varying coefficient vector
\begin{equation}
  \begin{bmatrix}
    a_0(t)  \\
    a_S(t)  \\
    a_A(t) 
  \end{bmatrix}  
   =  \exp{(i\omega_0t)} 
  \begin{bmatrix}
    c_0(t)  \\
    c_S(t)  \\
    c_A(t) 
  \end{bmatrix},  
\end{equation}
\eqref{tdse} leads to
\begin{equation}\label{tdse-2}
\frac{d}{dt}
  \begin{bmatrix}
    a_0(t)  \\
    a_S(t)  \\
    a_A(t) 
  \end{bmatrix}  
 = -i 
{\boldsymbol W}
  \begin{bmatrix}
    a_0(t)  \\
    a_S(t)  \\
    a_A(t) 
  \end{bmatrix},  
\end{equation}
where
\begin{equation}
\boldsymbol{W} = 
  \begin{bmatrix}
    0 & \alpha & \beta  \\
    \alpha & 0 & 0 \\
    \beta  &  0    & 0
  \end{bmatrix}.   
\end{equation}
The solution of \eqref{tdse-2} is thus
\begin{equation}\label{propagator0}
  \begin{bmatrix}
    a_0(t)  \\
    a_S(t)  \\
    a_A(t) 
  \end{bmatrix}  
 = \exp \left( -i \boldsymbol{W} 
  t \right)
  \begin{bmatrix}
    a_0(0)  \\
    a_S(0)  \\
    a_A(0) 
  \end{bmatrix}.  
\end{equation}

In the limit (assuming no plasmon dissipation, $\epsilon = 0$),
expanding the exponential and re-grouping terms, 
\eqref{propagator0} 
can be written more explicitly as
\begin{equation}\label{exact}
  \begin{array}{l}
  \begin{bmatrix}
    a_0(t)  \\
    a_S(t)  \\
    a_A(t) 
  \end{bmatrix}  
 = 
  \begin{bmatrix}
    a_0(0)  \\
    a_S(0)  \\
    a_A(0) 
  \end{bmatrix}  
+\\
\left(
  \begin{bmatrix}
    \eta^2 & 0 & 0  \\
    0 & \alpha^2 & \alpha \beta \\
    0 & \alpha \beta    & \beta^2 
  \end{bmatrix}   
\frac{F(t)}{\eta^2} -i \boldsymbol{W}
\frac{G(t)}{\eta}
   \right)
  \begin{bmatrix}
    a_0(0)  \\
    a_S(0)  \\
    a_A(0) 
  \end{bmatrix},  
\end{array}
\end{equation}
where
\begin{equation}
\eta = \sqrt {\alpha^2+\beta^2}
\end{equation}
and
\begin{equation}\label{FG}
F(t) = \cos (\eta t) - 1, \quad G(t) = \sin (\eta t).
\end{equation}

Note that for the initial condition
corresponding to $\ket{\Psi(t=0)} = \ket{q_2=0,q_1=0;s=0}$
or $a_0(0)=0$, $a_S(0) = a_A(0) = \frac{1}{\sqrt{2}}$,
the above exact solution (for $\epsilon = 0$) is such that 
$a_0(t)$ is a purely imaginary number for all times and 
that $a_S(t)$ and
$a_A(t)$ are purely real numbers for all times.
Equation \eqref{exact}, can be approximated to
various orders in time by expanding $F(t)$ and $G(t)$ defined
in \eqref{FG} appropriately. Thus, with the initial condition 
$\ket{\Psi(t=0)} = \ket{q_2=0,q_1=1;s=0}$,  
the approximate solution,
accurate to second order in time, is
\begin{equation}\label{approx}
  \begin{bmatrix}
    a_0(t)  \\
    a_S(t)  \\
    a_A(t) 
  \end{bmatrix}  
 \approx 
  \begin{bmatrix}
    -i\frac{(\alpha+\beta)t}{\sqrt{2}}   \\
    \frac{1}{\sqrt{2}}-\frac{(\alpha^2+\alpha \beta )t^2}{2\sqrt{2}}  \\
    \frac{1}{\sqrt{2}}-\frac{(\beta^2+\alpha \beta )t^2}{2\sqrt{2}}
  \end{bmatrix},
\end{equation}
or, in terms of $g_1$ and $g_2$,
\begin{equation}\label{approx_g}
  \begin{bmatrix}
    a_0(t)  \\
    a_S(t)  \\
    a_A(t) 
  \end{bmatrix}  
 \approx 
  \begin{bmatrix}
    -i g_1 t   \\
    \frac{1}{\sqrt{2}}-\frac{g_1 (g_1+g_2 )t^2}{2\sqrt{2}}  \\
    \frac{1}{\sqrt{2}}+\frac{g_1 (g_2-g_1 )t^2}{2\sqrt{2}}
  \end{bmatrix}.
\end{equation}

Of course, another way to obtain \eqref{exact}
is to determine the eigenvalues and eigenvectors of $\hat{W}$,
$w_k$ and $\ket{\phi_k}$, $k = 1,2,3$,
and represent \eqref{propagator0} with them. 
This procedure can be carried out exactly even when plasmonic dissipation
is allowed ($\epsilon > 0$).
The 
eigenvalues of $\boldsymbol{W}$ are easily found to be 
\begin{equation}\label{eigvals}
\begin{aligned}
w_1 &= 0 \\
w_2 &= \frac{1}{2} ( -i \epsilon - \sqrt{4 \alpha^2 + 4 \beta^2 - \epsilon^2}) \\
w_3 &= \frac{1}{2} ( -i \epsilon + \sqrt{4 \alpha^2 + 4 \beta^2 - \epsilon^2}),
\end{aligned}
\end{equation}
and the associated (unnormalized) eigenvectors projected onto the zero-order basis are
\begin{equation}\label{phi1}
  \begin{bmatrix}
    \braket{0,0;1 | \phi_1}  \\
    \braket{S;0 | \phi_1}  \\
    \braket{A;0 | \phi_1} 
  \end{bmatrix}  
 = 
  \begin{bmatrix}
    0  \\
    -\beta / \alpha  \\
    1 
  \end{bmatrix},  
\end{equation}
\begin{equation}\label{phi2}
  \begin{bmatrix}
    \braket{0,0;1 | \phi_2}  \\
    \braket{S;0 | \phi_2}  \\
    \braket{A;0 | \phi_2} 
  \end{bmatrix}  
 = 
  \begin{bmatrix}
\frac{-i \epsilon - \sqrt{4 \alpha^2 + 4 \beta^2 - \epsilon^2}}{2 \beta}      \\
    \alpha / \beta  \\
    1 
  \end{bmatrix},  
\end{equation}
and
\begin{equation}\label{phi3}
  \begin{bmatrix}
    \braket{0,0;1 | \phi_3}  \\
    \braket{S;0 | \phi_3}  \\
    \braket{A;0 | \phi_3} 
  \end{bmatrix}  
 = 
  \begin{bmatrix}
\frac{-i \epsilon + \sqrt{4 \alpha^2 + 4 \beta^2 - \epsilon^2}}{2 \beta}      \\
    \alpha / \beta  \\
    1 
  \end{bmatrix}.  
\end{equation}
The propagator may then be written as
\begin{equation}\label{propagator}
\exp (-i \hat{W}t) = 
\sum_k \ket{\phi_k} \bra{\phi_k^*} \exp (-i w_k t) / n_k,
\end{equation}
where 
\begin{equation}\label{norm}
n_k = \braket{\phi_k^* | \phi_k} = \sum_{j=0,S,A} \braket{j | \phi_k}^2.
\end{equation}
The bra vectors we employ, $\bra{c}$ (as is most common), are defined to be 
the transpose of the complex conjugates of the coefficients representing
their corresponding kets, $\ket{c}$.  
Thus $\braket{c | d} = \sum_j c_j^{*} d_j$, where $c_j = \braket{j|c}$,
$d_j = \braket{j|d}$.  
An expression such as
\eqref{norm}, which involves an additional complex conjugate in the
argument of the bra vector, implies that $n_k$ is the sum of the (complex) squares
of the  components of $\ket{\phi_k}$, as opposed to being the 
more familiar sum of the squares of
the magnitudes of the components.  
This necessary peculiarity arises from $\bf{W}$ being
symmetric but not Hermitian.  (In particular the symmetry of $\bf{W}$
implies
for eigenvalues  $w_a \neq w_b$ that $\braket{\phi_a^* | \phi_b} = 0$, which
ultimately leads to an expression for the 
unity operator involving a sum of
$\ket{\phi_k} \bra{\phi_k^*}$ terms instead of the more familiar sum of
$\ket{\phi_k} \bra{\phi_k}$ terms.)

We note that for $\epsilon = \gamma_s/2 > 0$, $w_2$ and $w_3$ {\em always}
have negative imaginary components. As 
$t \rightarrow \infty$, only the
$k = 1$ contribution to \eqref{propagator} survives because only $w_1$ has
no decay (or negative imaginary) component.  If we initiate the system with
one QD excited, then 
\begin{equation}\label{psi0}
  \begin{bmatrix}
    a_0(0)  \\
    a_S(0)  \\
    a_A(0) 
  \end{bmatrix}  
=
  \begin{bmatrix}
    0  \\
    \frac{1}{\sqrt{2}}  \\
    \frac{1}{\sqrt{2}} 
  \end{bmatrix}.  
\end{equation}
The asymptotic amplitude for $a_S$ is  then
\begin{equation}
\begin{aligned}
a_S ( \infty ) &= \braket{S| \Psi ( \infty )} \\
             &= \braket{S| \phi_1} {\braket{\phi_1 | \Psi (0)} } \\
&=  \frac{1}{\sqrt{2} (1+x^2)}
x (1 - x ),
\end{aligned}
\end{equation}
where 
\begin{equation}\label{x}
x = \frac{\beta}{\alpha} = \frac{g_1-g_2}{g_1+g_2},
\end{equation}
and we have used the fact that $n_1 = {1+x^2}$.
In a similar fashion we find
\begin{equation}
a_A ( \infty ) =  \frac{1}{\sqrt{2}(1+x^2)} (1 - x).
\end{equation}
The asymptotic concurrence in this case is simply~\cite{vidal-prb-2011} 
\begin{equation}
\begin{aligned}
C( \infty ) &=  \left|~ P_A ( \infty ) - P_S ( \infty )~ \right| \\
          &=    \left|~ | a_A( \infty ) |^2 ~ - ~| a_S (\infty) |^2 ~\right|.
\end{aligned}
\end{equation}
Since the magnitude of $x$ in \eqref{x}  is always less 
than 1 when $g_1$ and $g_2$ are positive,
$P_A > P_S$, and one can ignore the outer absolute signs.  
The concurrence then reduces to  
\begin{equation}\label{concurrence-f}
C( \infty ) =   \frac{1}{2(1+x^2)^2} ( 1 - x )^2  (1 - x^2).  
\end{equation}
Viewed as a function of $x$, the maximum of
\eqref{concurrence-f} is 
found to be at $x = -2 + \sqrt{3}$, and corresponds to $g_2/g_1 = \sqrt{3}$,
consistent with the results in the text.
For this value of $x$, $C ( \infty ) \approx 0.6495$.

The three-state model above involving the states $\ket{q_1=0,q_2=0;s=1}$, $\ket{S;s=0}$,
and $\ket{A;s=0}$ is convenient because it led directly to simple analytical 
expressions for
the asymptotic concurrence.  However, the same result can be
obtained, with a little more work, by employing the basis $\ket{q_1=0,q_2=0;s=1}$,
$\ket{q_1=0,q_2=1;s=0}$ and $\ket{q_1=1,q_2=0;s=0}$.  In fact this approach is advantageous
because it then is easily generalizable to $N > 2$ QDs.
Assume we have $N$ QDs, with each QD$k$ interacting only with the 
dissipative plasmon via a Hamiltonian coupling term $\hbar g_k$.
If the basis is taken to be $\ket{q_1=0,q_2=0,q_3=0,\ldots; s = 1}$,
$\ket{q_1=1,q_2=0,q_3=0,\ldots; s= 0}$, $\ket{q_1=0,q_2=1,q_3=0,\ldots;
s=0}$, \ldots,
then
one has an $(N+1) \times (N+1)$ Hamiltonian matrix representation 
$\bold{H} = \hbar \bold{W}$
with 
\begin{equation}\label{bigham}
{\bf W} = 
                \left[ 
                      \begin{array}{ccccc}
                             -i \epsilon &g_1&g_2&\cdots &g_N\\
                             g_1&0&0&\cdots &0\\
                             g_2&0&0&\cdots &0\\
                             \vdots&&&\ddots&\\
                             g_N&0&0&\cdots &0
                      \end{array}
                \right].
\end{equation}

The characteristic equation for the eigenvalues of ${\bf W}$ is
then $w^{N-1} (w^2 +i \epsilon w - G) = 0$, where
$G = \sum_{k=1}^N g_k^2$.  It implies that  there are $N-1$ 
degenerate eigenvalues
$w_1 = w_2 = \ldots = w_{N-1} = 0$ and two complex eigenvalues,
\begin{equation}
\begin{aligned}
w_{N}        &= (-i \epsilon - \sqrt{4G- \epsilon^2})/2 \\ 
w_{N+1}      &= (-i \epsilon + \sqrt{4G- \epsilon^2})/2.
\end{aligned}
\end{equation}

Let $\bf{v}^k$ denote the eigenvector corresponding
to the $k$th eigenvalue, and let $v_j^k$ denote the $j$th component of 
this eigenvector.  One can easily see that the $k = 1,2,\ldots, N-1$ 
degenerate eigenvectors 
must all have  $v_1^k = 0$; that is, they contain no component in the basis state
$\ket{q_1=0,q_2=0,\ldots;s=1}$.  The remaining components must satisfy
\begin{equation}\label{condition}
\sum_{j=2}^{N+1} g_{j-1} v_{j}^k = 0.
\end{equation}
Although one can easily solve \eqref{condition} for low $N$ in 
various ways, a systematic
procedure for obtaining $N-1$ linearly independent and orthogonal eigenvectors
is as follows.  Notice that \eqref{condition} implies that 
each of the desired vectors $\bf{v}^k$ must be orthogonal to the vector
$\bf{g}$ = $(0,g_1,\ldots,g_N)^T$.  Thus one can initially set $N-1$ vectors
with random coefficients and use a Gram-Schmidt procedure initiated
with the vector $\bf{g}$, orthogonalizing all subsequent vectors against
$\bf{g}$ and previously generated vectors.

The final two eigenvectors for $k = N$ and $k = N+1$ are easily found to have the $j = 1$ components
$v_1^N = w_N/g_N$ and $v_1^{N+1} = w_{N+1}/g_N$.  Their $j = 2,\ldots,N$
components are $v_j^{N} = v_j^{N+1} = g_{j-1}/g_N$ and, finally, for
the $j = N+1$ components, $v_{N+1}^N = v_{N+1}^{N+1} = 1$.
These two eigenvectors are orthogonal to each other and the previous $N-1$ 
eigenvectors associated with the degenerate eigenvalue, and we find it
convenient to employ them in this way with normalization considerations entering
into the propagator representation, \eqref{propagator}.

With the systematic procedure above for evaluating all the eigenvectors, and 
introducing
the time-dependent amplitudes $b_j(t)$ corresponding 
to states $j = 1,2,\ldots,N+1$ within the 
basis $\ket{0,0,0,0,\ldots;1}$, $\ket{0,1,0,0,\ldots;0}$, \ldots, $\ket{0,\ldots,0,1;0}$,
one can use \eqref{propagator} (extended to $N+1$ states, of course) to
show
\begin{equation}\label{final}
b_j(t) = \sum_{k=1}^{N+1} \exp (-i w_k t) K_{j,k},
\end{equation} 
where
\begin{equation}
K_{j,k} = \sum_i  v_j^k v_i^k b_i(0) /n_k. 
\end{equation} 

The probabilities for QDs $1,2,\ldots,N$ to be excited are $P_1 = \|b_2\|^2$,
$P_2 = \|b_3\|^2$, $\ldots,P_N = \|b_{N+1}\|^2$. 
While obtaining bipartite concurrences may appear
arduous, if $b_1(0) = 0$ (i.e., no amplitude in the state corresponds to the plasmon
excited with all QDs cold) and all 
the other amplitudes are real, one can show
that the bipartite concurrences are simply $C_{i,j} = 2 \sqrt{P_i P_j}$. 
As with the three-state example, we note that as $t
\rightarrow \infty$, only the $k = 1,2,\ldots,N-1$ eigenvector contributions
survive 
and one could use \eqref{final}, setting the exponential to one and carrying the
sum out to only $k = N-1$, to evaluate the asymptotic populations.

\section{Local Field Enhancement}\label{local-field-enhancement}
To estimate Rabi-flop frequencies for the QDs, we need an
estimate of the local electromagnetic field they experience, which is
enhanced relative to the incident field due to the presence of the 
plasmonic system.
To this end, we consider the interaction of one QD with a plasmonic
system and employ a classical coupled dipole picture, 
as in Ref.~\onlinecite{shah-prb-2013} and associated
supplementary material.  The time-dependent dipoles for the 
plasmon ($\mu_s(t)$) and
QD ($\mu_q(t)$), in the presence of an incident field with frequency
$\omega$ satisfy the equations of motion
\begin{equation}\label{mus-eom}
\ddot{\mu_s}(t) + \omega_s^2 \mu_s(t) + \gamma_s \dot{\mu_s}(t) = A_s [ E_0 
\cos \omega t + \mu_q(t) J ]
\end{equation}
\begin{equation}\label{muq-eom}
\ddot{\mu_q}(t) + \omega_q^2 \mu_q(t) + \gamma_q \dot{\mu_s}(t) = A_q [ E_0 
\cos \omega t + \mu_s(t) J ].
\end{equation} 
The parameters $\omega_s$, $\omega_q$, and $\gamma_s$ are the same
as those in the CQED model of \secref{sec:theoretical_methods}.
The other parameters in these classical equations are 
related to those in the
CQED model as follows: 
\begin{equation}\label{cl-to-cqed}
\begin{aligned}
J &= \frac{\hbar g}{d_s d_q} \\
A_s &= 2 d_s^2 \omega_s / \hbar \\
A_q &= 2 d_q^2 \omega_q / \hbar \\
\gamma_q &= 2 \gamma_d.
\end{aligned}
\end{equation}
Several comments are in order regarding these relations.
The relation for $J$ was derived in Ref.~\onlinecite{shah-prb-2013}. 
The relations for $A_s$ and $A_q$ reflect exactly solving \eqref{mus-eom} and
\eqref{muq-eom} in the limit of the dipoles not
interacting ($J = 0$) 
and equating the resulting
amplitudes of oscillation of the dipoles with the corresponding
quantum expressions (in the linear or low $E_0$ limit).  
These expressions are twice as small as the previously
inferred ones, which were less accurate because they were based
on an approximate solution of the classical equations.  
The classical decay factor $\gamma_q$ is taken to be twice the
corresponding quantum dephasing factor, $\gamma_d$.  This ensures
that the full-width-at-half-maximum of the isolated
QD spectrum, inferred from the classical expression with
$\gamma_q$, is equal to the corresponding quantum result in the low $E_0$ limit.

We can identify the  term $\mu_s(t)  J$ 
in 
\eqref{muq-eom} as the local electric field the QD experiences
because of the plasmon, that is,
\begin{equation} \label{elocal}
E^{loc}(t) = \mu_s(t) J.
\end{equation}  
For estimating $E^{loc}$, one can
approximate $\mu_s(t)$ by the 
expression
that results from  the exact solution of \eqref{mus-eom} in the
uncoupled ($J = 0$)  and on resonance ($\omega = \omega_s$) limits. This
solution is readily obtained by
complexifying the equation, that is, by replacing $\cos (\omega t)$ by 
$\exp (- i \omega t)$, which leads to an equation that is
easy to solve exactly.  The real part of the complex solution 
then solves the 
original, real equation. Thus,
\begin{equation} \label{mus}
\mu_s(t) \approx \frac{A_s E_0}{\omega_s \gamma_s} \sin ( \omega t ).
\end{equation}
Insertion of \eqref{mus} into \eqref{elocal} leads to
\begin{equation}
E^{\rm{loc}}(t) \approx E_0^{\rm{loc}} \sin (\omega t),
\end{equation}
where
\begin{equation}
E_0^{\rm{loc}} = 
2 \frac{d_s}{d_q}
\frac{g}{\gamma_s} E_0,
\end{equation}
where the expressions in \eqref{cl-to-cqed} have also been used.

% N.B. can include other bib files than prb, separated by commas. 

% use with separate report.bib; generates report.bbl which
% dumbo Phys. Rev. and Archive people prefer
%\bibliography{report}

\end{document}